\documentclass{emulateapj}

\usepackage{url}
\usepackage{epstopdf}
\usepackage{multirow}
\usepackage{amsmath}
\usepackage{natbib}
\usepackage{color}
\usepackage{booktabs}
\usepackage{tabularx}
\usepackage[driverfallback=dvipdfmx, breaklinks, colorlinks, citecolor=blue, linkcolor=blue, menucolor=blue, urlcolor=blue]{hyperref}
\usepackage{longtable}
\usepackage{rotating}

\slugcomment{}

\begin{document}

\title{Kinematics of the M87 jet in the collimation zone: gradual acceleration and velocity stratification}

\author{Jongho Park\altaffilmark{1,2},
Kazuhiro Hada\altaffilmark{3,4}, 
Motoki Kino\altaffilmark{5,6}, 
Masanori Nakamura\altaffilmark{1}, 
Jeffrey Hodgson\altaffilmark{7},
Hyunwook Ro\altaffilmark{8}, 
Yuzhu Cui\altaffilmark{3,4},
Keiichi Asada\altaffilmark{1},
Juan-Carlos Algaba\altaffilmark{2,7,9},
Satoko Sawada-Satoh\altaffilmark{10},
Sang-Sung Lee\altaffilmark{7,11},
Ilje Cho\altaffilmark{7,11},
Zhiqiang Shen\altaffilmark{12,13},
Wu Jiang\altaffilmark{12,13},
Sascha Trippe\altaffilmark{2}, 
Kotaro Niinuma\altaffilmark{10},
Bong Won Sohn\altaffilmark{7,8,11},
Taehyun Jung\altaffilmark{7,11},
Guang-Yao Zhao\altaffilmark{7},
Kiyoaki Wajima\altaffilmark{7},
Fumie Tazaki\altaffilmark{3},
Mareki Honma\altaffilmark{3,4},
Tao An\altaffilmark{12,13},
Kazunori Akiyama\altaffilmark{14},
Do-Young Byun\altaffilmark{7,8,11},
Jongsoo Kim\altaffilmark{7},
Yingkang Zhang\altaffilmark{12,15},
Xiaopeng Cheng\altaffilmark{12,15},
Hideyuki Kobayashi\altaffilmark{3},
Katsunori M. Shibata\altaffilmark{3,4},
Jee Won Lee\altaffilmark{7,16},
Duk-Gyoo Roh\altaffilmark{7},
Se-Jin Oh\altaffilmark{7},
Jae-Hwan Yeom\altaffilmark{7},
Dong-Kyu Jung\altaffilmark{7},
Chungsik Oh\altaffilmark{7},
Hyo-Ryoung Kim\altaffilmark{7},
Ju-Yeon Hwang\altaffilmark{7}, and
Yoshiaki Hagiwara\altaffilmark{17}
}

\affil{
$^1$ Institute of Astronomy and Astrophysics, Academia Sinica, P.O. Box 23-141, Taipei 10617, Taiwan; jpark@asiaa.sinica.edu.tw\\
$^2$Department of Physics and Astronomy, Seoul National University, Gwanak-gu, Seoul 08826, Republic of Korea\\
$^3$Mizusawa VLBI Observatory, National Astronomical Observatory of Japan, 2-21-1 Osawa, Mitaka, Tokyo 181-8588, Japan\\
$^4$Department of Astronomical Science, The Graduate University for Advanced Studies (SOKENDAI), 2-21-1 Osawa, Mitaka, Tokyo 181-8588, Japan\\
$^5$National Astronomical Observatory of Japan, 2-21-1 Osawa, Mitaka, Tokyo, 181-8588, Japan\\
$^6$Kogakuin University of Technology \& Engineering, Academic Support Center, 2665-1 Nakano, Hachioji, Tokyo 192-0015, Japan\\
$^7$Korea Astronomy and Space Science Institute, Yuseong-gu, Daejeon 34055, Korea\\
$^8$Department of Astronomy, Yonsei University, 134 Shinchondong, Seodaemungu, Seoul 120-749, Republic of Korea\\
$^{9}$Department of Physics, Faculty of Science, University of Malaya, 50603 Kuala Lumpur, Malaysia\\
$^{10}$Graduate School of Sciences and Technology for Innovation, Yamaguchi University, Yoshida 1677-1, Yamaguchi, Yamaguchi 753-8512, Japan\\
$^{11}$Department of Astronomy \& Space Science, University of Science \& Technology, 217 Gajeong-ro, Daejeon, Republic of Korea\\
$^{12}$Shanghai Astronomical Observatory, Chinese Academy of Sciences, Shanghai 200030, People’s Republic of China\\
$^{13}$Key Laboratory of Radio Astronomy, Chinese Academy of Sciences, 210008 Nanjing, People’s Republic of China\\
$^{14}$Massachusetts Institute of Technology, Haystack Observatory, 99 Millstone Road, Westford, MA 01886, USA\\
$^{15}$University of Chinese Academy of Sciences, 19A Yuquanlu, Beijing 100049, People’s Republic of China\\
$^{16}$Department of Astronomy and Space Science, Kyung Hee University, 1732, Deogyeong-daero, Giheung-gu, Yongin-si, Gyeonggi-do 17104, Korea\\
$^{17}$Toyo University, 5-28-20 Hakusan, Bunkyo-ku, Tokyo 112-8606, Japan \\
}

\received{...}
\accepted{...}

\begin{abstract}
\noindent We study the kinematics of the M87 jet using the first year data of the KVN and VERA Array (KaVA) large program, which has densely monitored the jet at 22 and 43 GHz since 2016. We find that the apparent jet speeds generally increase from $\approx0.3c$ at $\approx0.5$ mas from the jet base to $\approx2.7c$ at $\approx20$ mas, indicating that the jet is accelerated from subluminal to superluminal speeds on these scales. We perform a complementary jet kinematic analysis by using archival Very Long Baseline Array monitoring data observed in $2005-2009$ at 1.7 GHz and find that the jet is moving at relativistic speeds up to $\approx5.8c$ at distances of $200-410$ mas. We combine the two kinematic results and find that the jet is gradually accelerated over a broad distance range that coincides with the jet collimation zone, implying that conversion of Poynting flux to kinetic energy flux takes place. If the jet emission consists of a single streamline, the observed trend of jet acceleration ($\Gamma\propto z^{0.16\pm0.01}$) is relatively slow compared to models of a highly magnetized jet. This indicates that Poynting flux conversion through the differential collimation of poloidal magnetic fields may be less efficient than expected. However, we find a non-negligible dispersion in the observed speeds for a given jet distance, making it difficult to describe the jet velocity field with a single power-law acceleration function. We discuss the possibility that the jet emission consists of multiple streamlines following different acceleration profiles, resulting in jet velocity stratification.

\end{abstract}


\section{Introduction \label{sect1}}

Active galactic nuclei (AGNs) often produce highly collimated relativistic jets (e.g., \citealt{Blandford2019}). Apparent superluminal motions are seen in these jets with speeds observed up to tens of times the speed of light ($c$, e.g., \citealt{Lister2016, Jorstad2017}). This indicates that the intrinsic bulk velocity is likely a large fraction of $c$. It is widely believed that the jets, after they are launched in the vicinity of the central supermassive black holes by the accretion of matter \citep{Meier2012}, are collimated and accelerated simultaneously at distances $\lesssim10^4-10^6\ R_{\rm S}$, where $R_{\rm S}$ is the Schwarzschild radius (e.g., \citealt{Meier2001, Marscher2008}). Very long baseline interferometry (VLBI) observations which provide a high angular resolution have been extensively used to study the jet acceleration and collimation mechanism (e.g., \citealt{Homan2015}).

\begin{deluxetable*}{ccccccc}[t!]

\tablecaption{Summary of KaVA observations in 2016}
\tablehead{
\colhead{Exp. Code} & \colhead{Obs. Date} &
\colhead{Stations} &
\colhead{Beam size (mas$\times$mas, deg)} & \colhead{$I_{\rm peak}$ (Jy/beam)} & \colhead{$I_{\rm rms}$ (mJy/beam)} & Dynamic Range \\
(1) & (2) & (3) & \centering(4) & (5) & (6) & (7)
}
\startdata
\multicolumn{7}{c}{22 GHz} \\
\midrule
 k16mk02a & 2016 Feb 25 (56d) & KaVA & $1.34\times1.20, -6.1$ & 1.41 & 0.42 & 3300 \\
 k16mk02c & 2016 Mar 09 (69d) & KaVA & $1.42\times1.20, -0.2$ & 1.42 & 0.50 & 2805 \\
 k16mk02e & 2016 Mar 21 (81d) & KaVA & $1.50\times1.25, -12.0$ & 1.45 & 0.51 & 2860 \\
 k16mk02g & 2016 Apr 08 (99d) & KaVA & $1.35\times1.16, -10.3$ & 1.29 & 0.30 & 4336 \\
 k16mk02i & 2016 Apr 21 (112d) & KaVA & $1.32\times1.23, -10.9$ & 1.28 & 0.36 & 3630 \\
 k16mk02k & 2016 May 03 (124d) & KaVA & $1.35\times1.10, -13.4$ & 1.17 & 0.32 & 3748 \\
 k16mk02m & 2016 May 23 (144d) & KaVA & $1.27\times1.10, -12.6$ & 1.15 & 0.32 & 3646 \\
 k16mk02q & 2016 Jun 13 (165d) & KaVA & $1.25\times1.13, -3.1$ & 1.21 & 0.68 & 1776\\
 \midrule
 \multicolumn{7}{c}{43 GHz} \\
\midrule
 k16mk02b & 2016 Feb 26 (57d) & KaVA & $0.76\times0.63, 16.2$ & 1.10 & 0.42 & 2624 \\
 k16mk02d & 2016 Mar 10 (70d) & KaVA & $0.79\times0.68, -0.24$ & 1.08 & 0.44 & 2405 \\
 k16mk02f & 2016 Mar 20 (80d) & KaVA & $0.72\times0.64, -30.3$ & 1.04 & 0.33 & 3255 \\
 k16mk02h & 2016 Apr 09 (100d) & KaVA & $0.71\times0.62, -7.5$ & 0.95 & 0.35 & 2665 \\
 k16mk02j & 2016 Apr 22 (113d) & KaVA & $0.70\times0.61, -26.0$ & 0.90 & 0.38 & 2390 \\
 k16mk02n & 2016 May 24 (145d) & KaVA & $0.64\times0.57, 2.1$ & 0.86 & 0.48 & 1794 \\
 k16mk02p & 2016 Jun 02 (154d) & KaVA, -MIZ & $0.83\times0.65, 43.1$ & 0.82 & 0.42 & 1978 \\
 k16mk02r & 2016 Jun 15 (167d) & KaVA & $0.71\times0.52, -34.0$ & 0.78 & 0.62 & 1240
\enddata
\tablecomments{(1) Experiment code of KaVA observations. (2) Observation date. Those in the parentheses denote the number of days elapsed since 2016 Jan 1. (3) Stations participating in observations. KaVA means that all seven stations successfully participated in observations. In the observation performed on 2016 Jun 02 at 43 GHz, the VERA Mizusawa station (MIZ) could not participate due to technical problems. (4) Full width at half maximum of the synthesized beam of M87 data with a natural weighting scheme. (5) Peak intensity of M87 with a natural weighting scheme. (6) Off-source rms noise of M87 maps with a natural weighting scheme. (7) Dynamic range of M87 images calculated from $I_{\rm peak}/I_{\rm rms}$. \label{Info}}
\end{deluxetable*}

M87 is the bright nucleus of the Virgo cluster and is a primary target for studies of AGN jets. The black hole shadow revealed by recent Event Horizon Telescope (EHT) observations \citep{EHT2019a, EHT2019b, EHT2019c, EHT2019d, EHT2019e, EHT2019f} confirms the existence of the supermassive black hole of mass of $M_{\rm BH} = (6.5\pm0.7)\times10^9M_\odot$ (\citealt{EHT2019f}, see also \citealt{Gebhardt2011}). It is located at a distance of 16.8 Mpc (\citealt{EHT2019c}; based on distance measurements of \citealt{Blakeslee2009, Bird2010, Cantiello2018}), giving a scale of 1 mas $\approx130\ R_{\rm S}$. Previous VLBI observations revealed that the jet shows an edge-brightened morphology and that the apparent jet opening angle becomes larger at smaller distances from the core on mas scales (up to $\gtrsim100^\circ$ at $\approx0.1$ mas from the core, e.g., \citealt{Reid1989, Junor1999, Hada2016, Kim2018, Walker2018}), indicating that the jet is being substantially collimated. The collimation continues at larger distances up to about 400 mas following a semi-parabolic profile of $R\propto z^{0.56}$, where $R$ and $z$ denote the jet radius and distance from the jet launching region, respectively \citep{AN2012, Hada2013, NA2013}. Recent general relativistic magnetohydrodynamic (GRMHD) simulations \citep{Nakamura2018} and a study of Faraday rotation in the jet \citep{Park2019} suggested that an external medium surrounding the jet, possibly non-relativistic winds launched from the accretion flows, may play a dynamical role in the jet collimation process (see also \citealt{Tseng2016, Boccardi2016, Boccardi2019, Algaba2017, Pushkarev2017, Akiyama2018, Giovannini2018, Hada2018, Nakahara2018, Nakahara2019a, Nakahara2019b, Kovalev2019} for a recent progress in jet collimation studies of other sources).

While the jet collimation profile is constrained precisely in a broad range of jet distances, the jet acceleration profile is under debate. Superluminal motions of several components at apparent speeds up to $\approx6.1c$ at optical wavelengths \citep{Biretta1999} and up to $\approx5.1c$ at radio wavelengths \citep{Cheung2007, Giroletti2012, Hada2015} are observed in a region known as HST-1, which is almost coincident with the Bondi radius \citep{Russell2015}. This region is located at an angular distance $\sim900$ mas from the core and has been a source of active study for decades (e.g., \citealt{Harris2003, Harris2009, Stawarz2006, Cheung2007, BL2009, Nakamura2010, Giroletti2012, NM2014, LG2017}). The jet apparent speeds become smaller beyond this location \citep{Biretta1995, Biretta1999, Meyer2013}, implying that the jet acceleration mostly occurs before HST-1. \cite{Kovalev2007} reported subluminal motions of several components within the distance of $\approx25$ mas from the monitoring of the jet over $\approx12$ years at 15 GHz with the Very Long Baseline Array (VLBA). Similar results of very slow or no apparent motions were obtained at lower observing frequencies \citep{Reid1989, Dodson2006}. \cite{Asada2014} found that the jet motions remain subluminal until $\approx200$ mas and the jet is substantially accelerated to relativistic speeds between $\sim200$ and $\sim400$ mas by using the data observed in three epochs in 2007--2009 at 1.6 GHz with the European VLBI Network. However, recent studies using densely-sampled data with the VLBA at 43 GHz \citep{Mertens2016, Walker2018} and with KaVA (KVN and VERA Array, \citealt{Niinuma2014, Zhao2019}) at 22 GHz \citep{Hada2017} have detected superluminal motions at distances $\lesssim20$ mas, which is in contradiction with the earlier studies. Additionally, these studies showed that the jet is substantially accelerated already at projected distances $\gtrsim0.5$ mas.

To resolve this discrepancy and to constrain the jet acceleration profile more accurately, we started a dedicated monitoring program of M87 in the framework of a KaVA Large Program in 2016 \citep{Kino2015a, Hada2017}. In this program, M87 is observed biweekly over four to seven months every year at both 22 and 43 GHz quasi-simultaneously. The good sensitivity (with a typical dynamic range of $\approx2000-4000$ for M87 at 22 GHz, \citealt{Hada2017}), the reasonably good angular resolution ($\approx1.2$ mas and $\approx 0.6$ mas at 22 and 43 GHz, respectively), and the good $uv$-coverage of KaVA especially for short baselines (Figure~\ref{uv}) make it possible to investigate the jet velocity field at various distances from the core; up to $\approx25$ mas at 22 GHz. KaVA has recently expanded to the East Asian VLBI Network (EAVN) which includes 21 telescopes in total and covers a wide range of observing frequencies from 2.3 to 43 GHz \citep{Wajima2016, Asada2017, Cho2017, An2018}. Our large program has been using the EAVN since 2017. In this paper, we report the results of the jet kinematics in M87 by using the KaVA-only observations performed in 2016. 

The paper is organized as follows. We describe the observations, the KaVA large program, and data reduction in Section~\ref{sect2}. We summarize the methods used for the M87 jet kinematics in previous studies in Section~\ref{sect3}. We present the results of jet kinematics obtained with KaVA observations in Section~\ref{sect4}. In Section~\ref{sect5}, we supplement our jet proper motion measurements with archival VLBA data observed in 2005--2009 at 1.7 GHz, which can trace the jet motion on larger scales compared to our KaVA data. We discuss the possible implications of our results in Section~\ref{sect6} and conclude in Section~\ref{sect7}. In this work, we adopt a jet viewing angle of $17^\circ$ \citep{Mertens2016, Walker2018}.

\section{Observations and Data Reduction}
\label{sect2}

\begin{figure}[t!]
\begin{center}
\includegraphics[trim=1mm 7mm 8mm 11mm, clip, width = 0.48\textwidth]{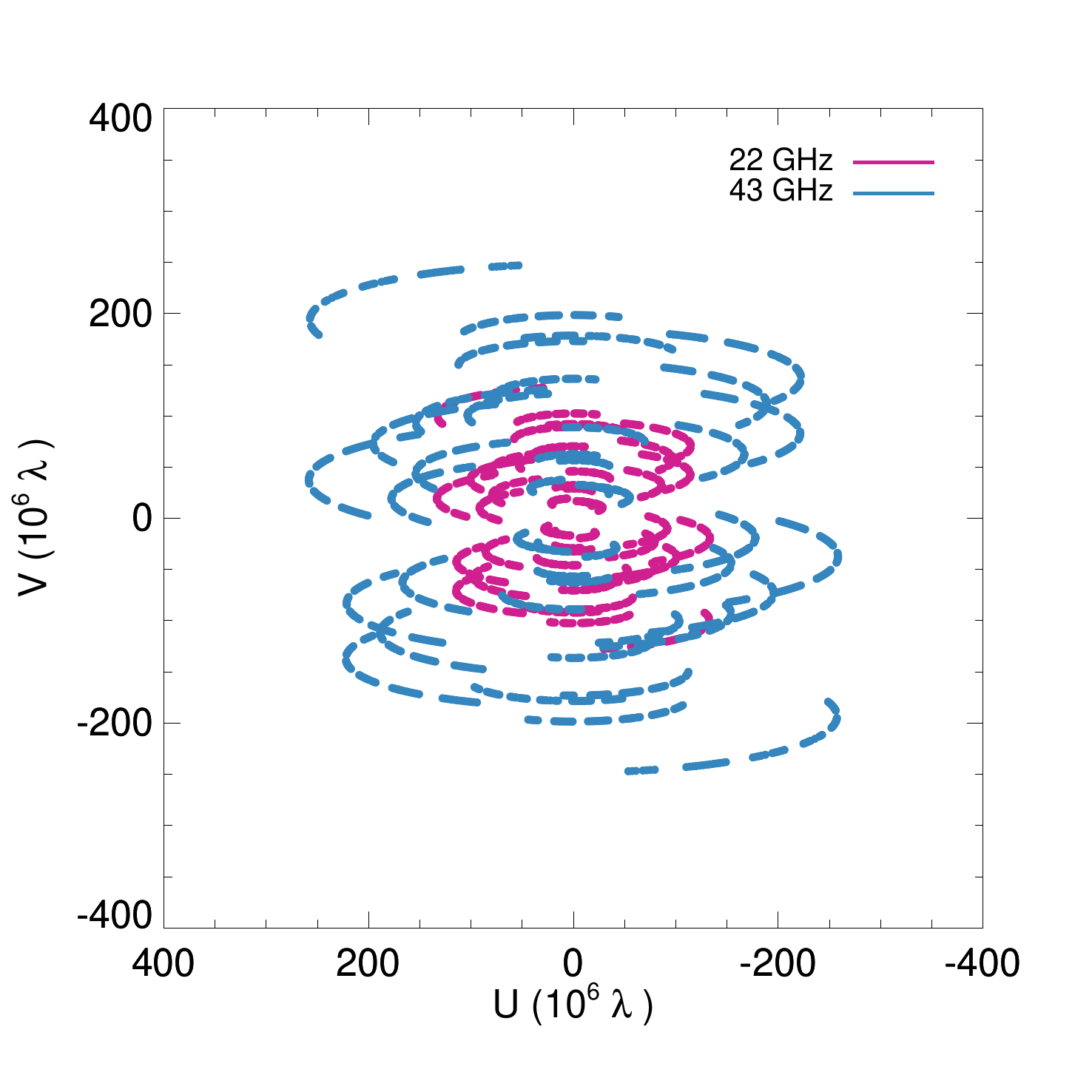}
\caption{Typical $uv$-coverage of KaVA observations of M87 taken from the first epoch data at 22 (1.3 cm, red) and 43 GHz (7 mm, blue) in units of $10^6$ times the wavelengths.
\label{uv}}
\end{center}
\end{figure}

We observed M87 with KaVA nine times in 2016. Observations in each epoch were performed in two sessions, one at 22 GHz and the other at 43 GHz, separated by one or two days. The monitoring interval between adjacent epochs is typically two weeks. The on-source time for M87 is about four and a half hours out of the total observing time of seven hours for each epoch at each frequency, allowing us to achieve a good $uv$-coverage (Figure~\ref{uv}). The typical beam size is about 1.2 and 0.6 mas at 22 and 43 GHz, respectively. Under the natural weighting of the visibility data, the beam shape is close to a circular shape, as seen in previous KaVA observations (e.g., \citealt{Niinuma2014, Oh2015, Hada2017, Lee2019}). All seven KaVA stations successfully participated in the observations in most epochs, except for two epochs. On 2016 Jun 01, we lost two VERA (Mizusawa and Ishigaki) stations at 22 GHz and we did not include this data in the current paper. On 2016 Jun 02, we lost the Mizusawa station at 43 GHz. The weather conditions were very good in general, providing us with a set of high-quality images with a typical dynamic range of 3000--4000 and 2000--3000 at 22 and 43 GHz, respectively. However, the data observed on 2016 May 05 at 43 GHz suffered from severe weather conditions at various stations, and we excluded this data from our analysis. Thus, we use the data observed in eight epochs in total at each frequency. We summarize the basic information of our observations in Table~\ref{Info}.

Our data were recorded in left-hand circular polarization with two-bit quantization in 8 IFs (baseband channels, each IF consists of 256 channels) at a recording rate of 1 Gbps, yielding a total bandwidth of 256 MHz, and correlated by the Daejeon correlator at the Korea-Japan Correlation Center (KJCC, \citealt{Lee2014, Lee2015a}). We performed a standard data post-correlation process with the NRAO's Astronomical Image Processing System (AIPS, \citealt{Greisen2003}). A priori amplitude calibration was performed by using the antenna gain curves and system temperatures. We obtained models of the amplitudes of the antenna bandpass shapes by using the auto-correlation data of scans on bright calibrators, with which we normalized the bandpass shapes of all data. We scaled up the amplitudes by a factor of 1.3 to account for the known (constant) amplitude loss of the Daejeon hardware correlator \citep{Lee2015b, Hada2017}. We corrected antenna parallactic angles for the three KVN antennas only because the field rotators in the receiving rooms of VERA antennas fix the parallactic angle during observations. We also corrected instrumental delays in the visibility phases by using scans on bright calibrators. We performed a global fringe fitting with a solution interval between 10 and 30s for each IF, depending on the weather conditions. The data were averaged over the channels within each IF, and we performed imaging with an iterative procedure of CLEAN and phase/amplitude self-calibration in the Caltech Difmap package \citep{Shepherd1997}. We present naturally weighted CLEAN images at 22 and 43 GHz in Figures~\ref{cleank} and~\ref{cleanq}, respectively.

\section{Summary of Previous Studies of the M87 Jet Kinematics}
\label{sect3}

One of the most important issues in determining the jet kinematics is how to identify each part of the jet in different epochs. Various methods have been employed to study the M87 jet kinematics. Each method has strengths and weaknesses, which we summarize below.

(i) {\tt Modelfit} with Gaussian components. This method fits several components of circular or elliptical Gaussian brightness distributions to the visibility data, describing the observed jet structure with several distinct regions. This is a standard method for determining the jet kinematics of radio-loud AGNs (e.g., \citealt{Lister2016, Jorstad2017}). One of the biggest advantages of this method is that it is easy to identify different components in different epochs, especially when the total number of components in different epochs is the same. However, the M87 jet shows a complex jet structure with a prominent limb-brightening (e.g., \citealt{Junor1999, Kovalev2007, Hada2016, Kim2018, Walker2018}), making it difficult to determine if the assumption of a simple Gaussian brightness distribution for each jet region can be applied to M87 or not. \cite{Kovalev2007} applied this method to their long-term monitoring data of the M87 jet obtained with the VLBA at 15 GHz and found apparent speeds $\lesssim0.05c$. \cite{Asada2014} obtained the jet acceleration profile between $\approx200$ and $\approx400$ mas, by making use of this method as well. \cite{Hada2017} also used this method for their KaVA monitoring data obtained in 2014 and derived relatively fast motions at apparent speeds of up to $\approx2c$ with an indication of jet acceleration at $\lesssim10$ mas. We note that they also `grouped' two components at distances of 3--6 mas and treated them as a single component to obtain the velocities because of the complicated component identification for this region. \cite{Britzen2017} obtained circular Gaussian components for the north and south jet edges separately for some epochs by using the MOJAVE program (\citealt{LH2005}, part of this data were presented in \citealt{Kovalev2007}), and obtained hints of jet acceleration at distances $\lesssim10$ mas.

(ii) Visual inspection. The complex jet structure and the complicated component identification led several studies to pick characteristic patterns in the jet brightness distributions by visual inspection in order to obtain the velocities. \cite{Ly2007} measured the apparent speeds of $0.25-0.4c$ at distances $\approx2-4$ mas from the three locally brightened positions forming a triangular shape in two epochs of 2001.78 and 2002.42 of their VLBA observations at 43 GHz. \cite{Hada2016} identified four and five locally brightened components along the north and south jet limbs, respectively, and one component in the counterjet, using their VLBA observations at 43 and 86 GHz. They fitted an elliptical Gaussian function to each component in the image plane to determine the component position and found the apparent speeds of $0.15-0.48c$. \cite{Walker2018} visually determined the locations of local maxima in the total intensity maps and identified components in different epochs by blinking rapidly back and forth between the maps in different epochs. They found an indication of jet acceleration at $\lesssim5$ mas for both north and south jet limbs with apparent speeds of up to $\approx5c$. Although this method is straightforward, the component identification would suffer from a lack of objectivity, especially when the total numbers of components in different epochs are not the same.

(iii) Subtracting the average image from the individual epoch images. \cite{Acciari2009} subtracted the average image of 11 epochs data observed in 2007 with the VLBA at 43 GHz from the individual epoch images in 2008, and they traced the bright regions in the subtracted images, obtaining an apparent speed of $1.1c$ at 0.77 mas. This method assumes that the brightness enhancement near the core (at $\lesssim0.5$ mas) is due to a new moving component ejected from the core. We note that they could use this method due to a significant increase in flux density of the inner jet observed in 2008 which was coincident with a flare seen at TeV energies. However, this method might not be applicable in a more generalized case. We also note that \cite{Walker2018} could not detect such a high apparent speed at the given distance, even if they used more data sets including those used in \cite{Acciari2009}.

(iv) Wavelet-based Image Segmentation and Evaluation (WISE). This method allows one to decompose and segment images and to identify significant structural patterns (SSPs) in different epochs through the multiscale cross-correlation (MCC) method, providing an automated or unsupervised way to obtain the jet velocity field \citep{ML2015}. \cite{Mertens2016} applied this method to the VLBA monitoring data of M87 observed in 11 epochs in 2007 at 43 GHz and revealed rich information about the velocity field at $\lesssim6$ mas with a clear indication of jet acceleration on these scales. They also applied the stacked cross-correlation algorithm \citep{ML2016} and found that there are at least two layers in the jet at 1--4 mas, one moving at a superluminal speed of $\gtrsim2c$ and the other at a subluminal speed of $\lesssim0.5c$. However, this method has not yet been verified on data sets having different $uv$-coverages, angular resolutions, imaging sensitivities, and sampling intervals.

(v) The brightness ratio of the jet and counterjet. Previous VLBI observations showed that there is tenuous but significant jet emission on the counterjet side (e.g., \citealt{Ly2007, Kovalev2007, Hada2016, Mertens2016, Walker2018, Kim2018}). The location of the jet base, probably coincident with the location of the black hole, was constrained to be quite close to the positions of the core at cm wavelengths, i.e., $\lesssim0.04$ mas from the 43 GHz core \citep{Hada2011}, indicating that the weak jet emission on the eastern side of the core is a counterjet. When assuming that the jet and the counterjet are intrinsically symmetric, and that there is no substantial free-free absorption towards the counterjet by the accretion flows (which seems to be the case for M87, see \citealt{Ly2007}, see also, e.g., \citealt{Jones1996, JW1997, Walker2000, FN2017} for other nearby radio galaxies), then the brightness ratio between the jet and counterjet at the same distance from the jet base can be explained by the result of Doppler boosting. Specifically, the brightness ratio is related to the intrinsic jet speed in units of the speed of light ($\beta$) as follows\footnote{We consider the case of a continuous jet for the beaming factor \citep{Ghisellini1993}.}:

\begin{equation}
\frac{I_{\rm jet}}{I_{\rm cjet}} = \left(\frac{1+\beta\cos\theta}{1-\beta\cos\theta}\right)^{2-\alpha},
\end{equation}

\noindent where $I_{\rm jet}$ and $I_{\rm cjet}$ denote the jet and counterjet intensity, respectively, $\alpha$ is the spectral index of the synchrotron radiation ($I_\nu \propto \nu^\alpha$), and $\theta$ is the jet viewing angle. The intrinsic jet speed can be converted into the apparent speed ($\beta_{\rm app}$) via

\begin{equation}
\beta_{\rm app} = \frac{\beta\sin\theta}{1 - \beta\cos\theta}.
\end{equation}

\noindent This method does not suffer from the complicated characterization and identification of jet `components' in different epochs. However, imaging the counterjet emission in VLBI observations is usually subject to relatively large calibration and deconvolution errors, which may introduce relatively large errors in the measured brightness ratio (e.g., \citealt{Ly2004}). Combining the measurements of the brightness ratio in different studies, using the adopted jet viewing angle of $17^\circ$ and the spectral index of $\alpha = -0.7\pm0.2$ for the inner jet region at 43 GHz \citep{Hada2016}, one can obtain the apparent speeds of $\sim0.1-0.4c$ at $\sim0.2-1.0$ mas \citep{Ly2007, Hada2016, Mertens2016, Walker2018, Kim2018}.

\section{Jet Kinematics on Scales of $\lesssim20$ mas Based on KaVA Observations}
\label{sect4}

Among the five methods of the jet kinematics listed above, we applied methods (i, Section~\ref{circularGaussian}), (ii, Section~\ref{point}), and (iv, Section~\ref{sectwise}) to our data. We could not find significant brightening of the core and the inner jet emission during the period of our observations and therefore we were not able to apply method (iii). Instead, the core intensity decreases with time (Table~\ref{Info}). This will be examined in a forthcoming paper by combining other data sets observed in different periods (Y. Cui et al. 2019, in preparation). Although we have found evidence of counterjet emission in our data, the limited angular resolution of KaVA makes it difficult to apply method (v). We describe how we obtain the kinematic results with each method below.

\begin{figure*}[t!]
\begin{center}
\includegraphics[trim=15mm 10mm 5mm 15mm, clip, width = 0.49\textwidth]{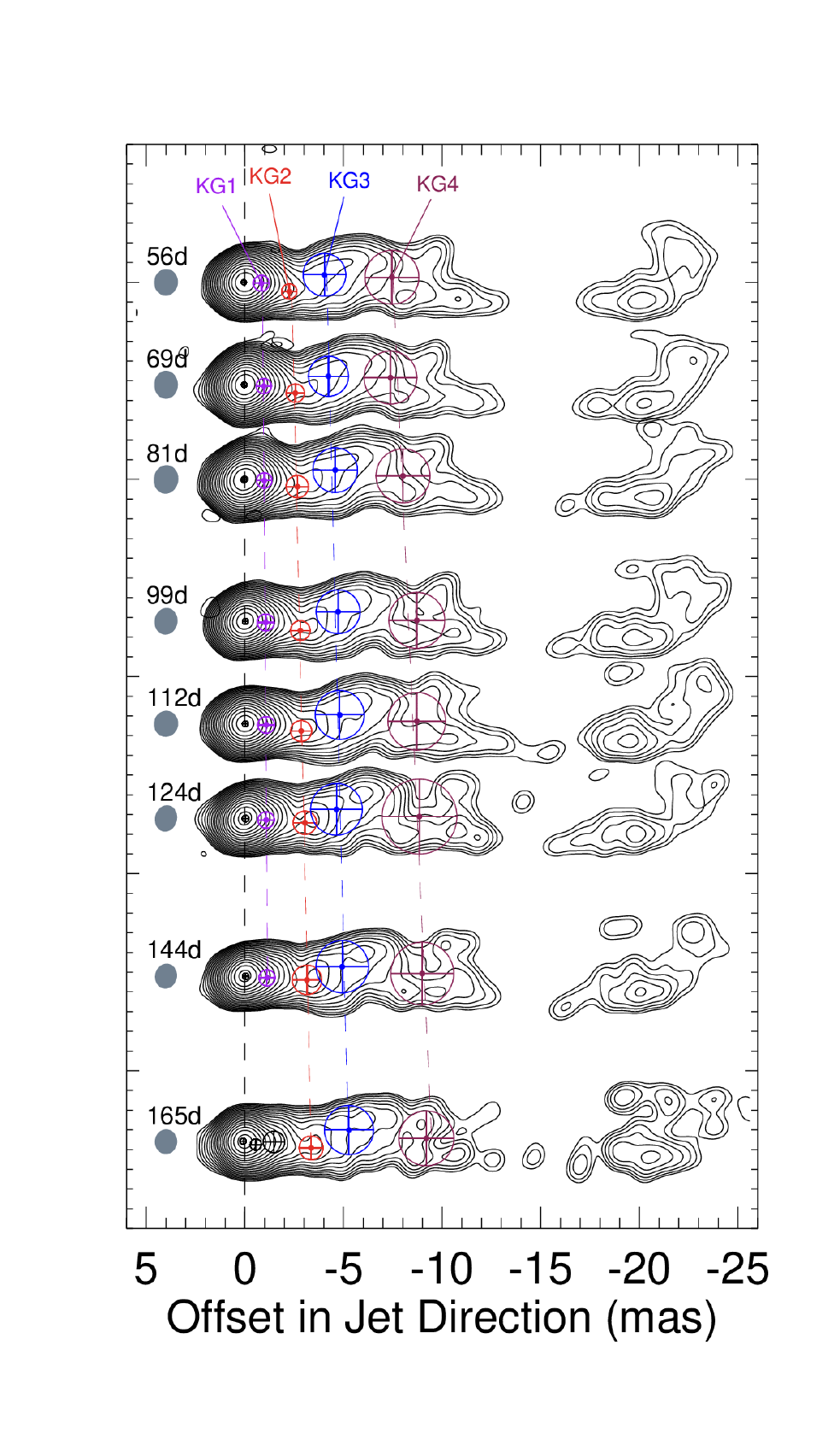}
\includegraphics[trim=15mm 10mm 5mm 15mm, clip, width = 0.49\textwidth]{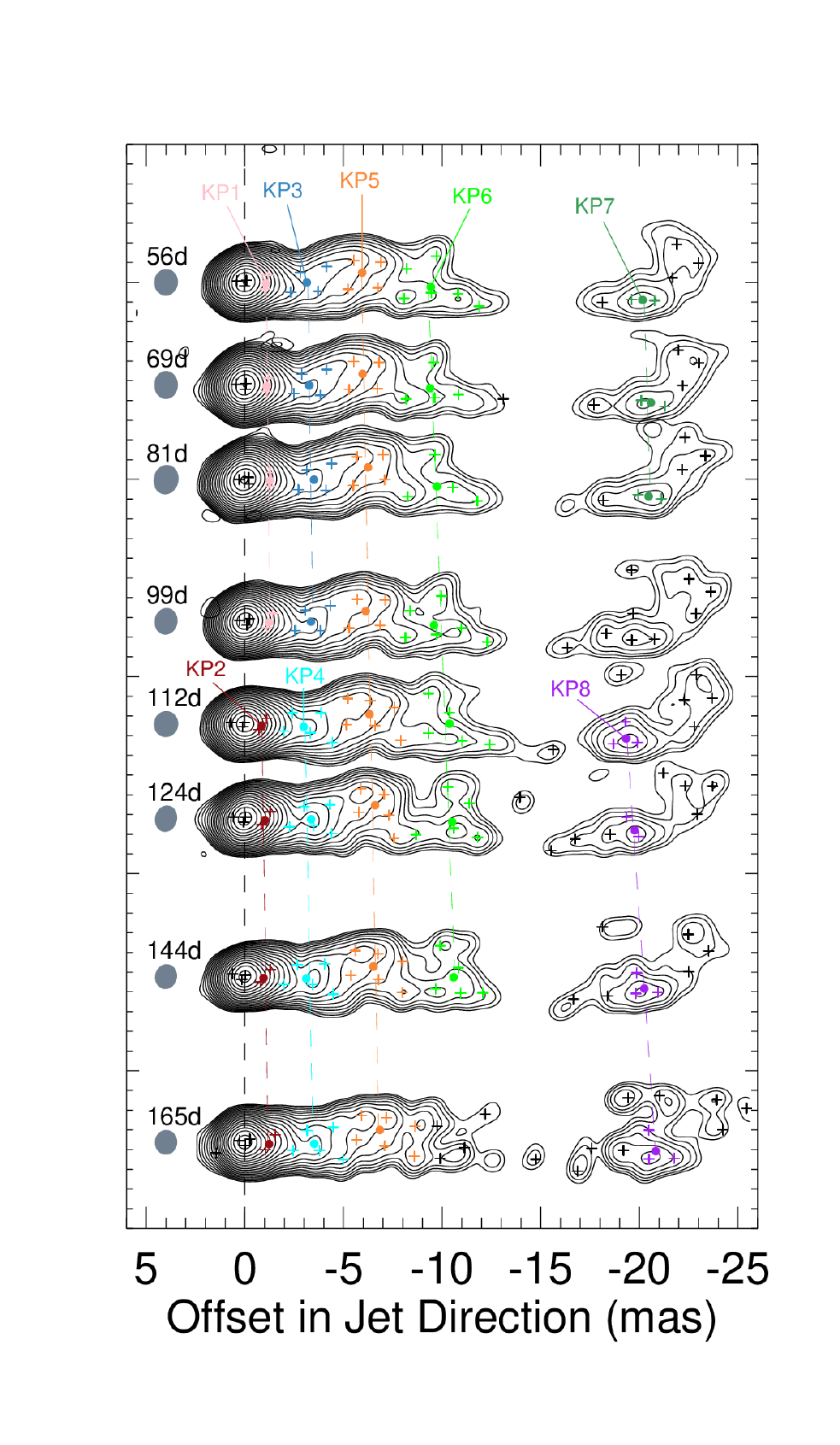}
\caption{Contours show CLEAN images of the M87 jet obtained with KaVA in 2016 at 22 GHz. The model components obtained in the {\tt modelfit} analysis with circular Gaussian components (Section~\ref{circularGaussian}) and with point source components (Section~\ref{point}) are drawn on top of the contours in the left and right panels, respectively. The components with the same color in different epochs are identified to represent the same parts of the jet. The point source models in the right panel are grouped and treated as a single component and their mean positions weighted by flux density are shown with the small filled circles. The vertical spacing is proportional to the time elapsed. The dashed lines show the best-fit linear motions of the components. The gray shaded ellipses at RA = 4 mas denote the full width half maximum of the synthesized beam. The number of days elapsed since 2016 Jan 1 is noted for each map. All maps are rotated clockwise by $18^\circ$ with respect to the map center in each epoch. Contours start at 1.9 mJy per beam and increase by factors of $\sqrt{2}$.
\label{cleank}}
\end{center}
\end{figure*}

\begin{figure*}[t!]
\begin{center}
\includegraphics[trim=13mm 23mm 8mm 24mm, clip, width = 0.49\textwidth]{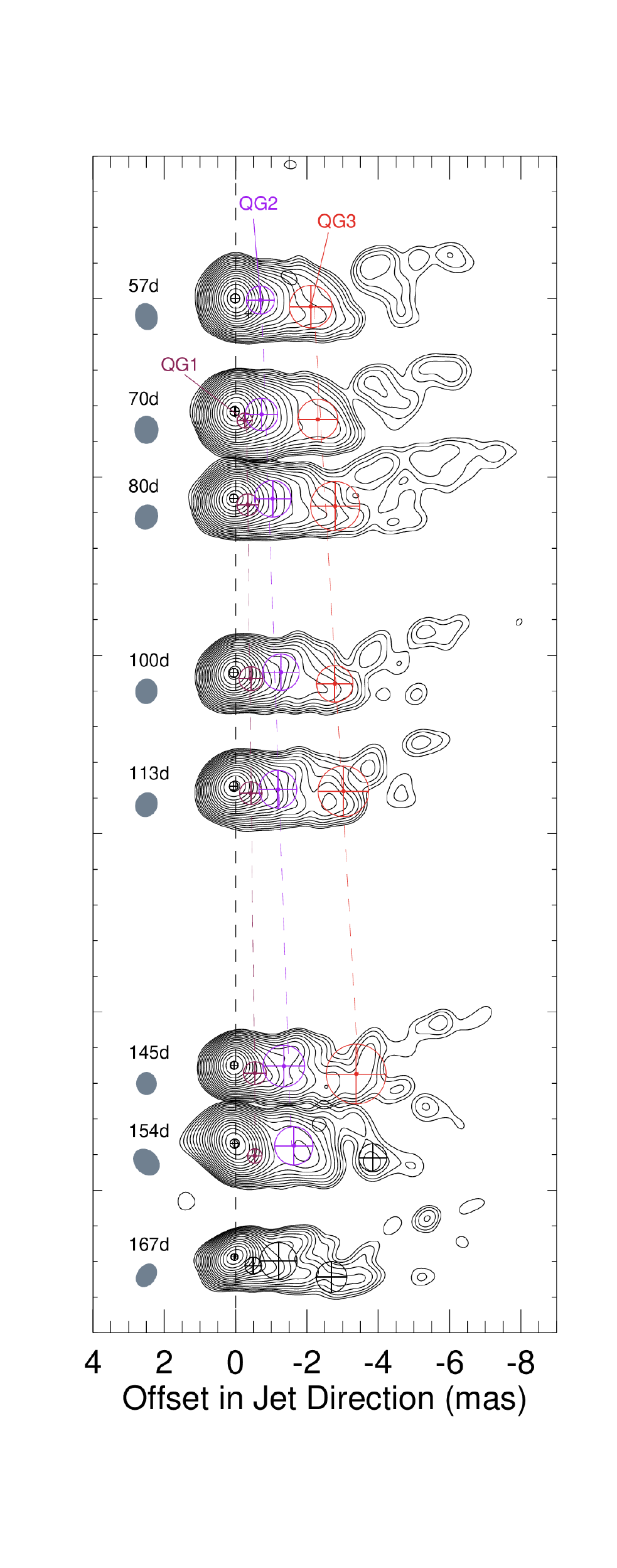}
\includegraphics[trim=13mm 23mm 8mm 24mm, clip, width = 0.49\textwidth]{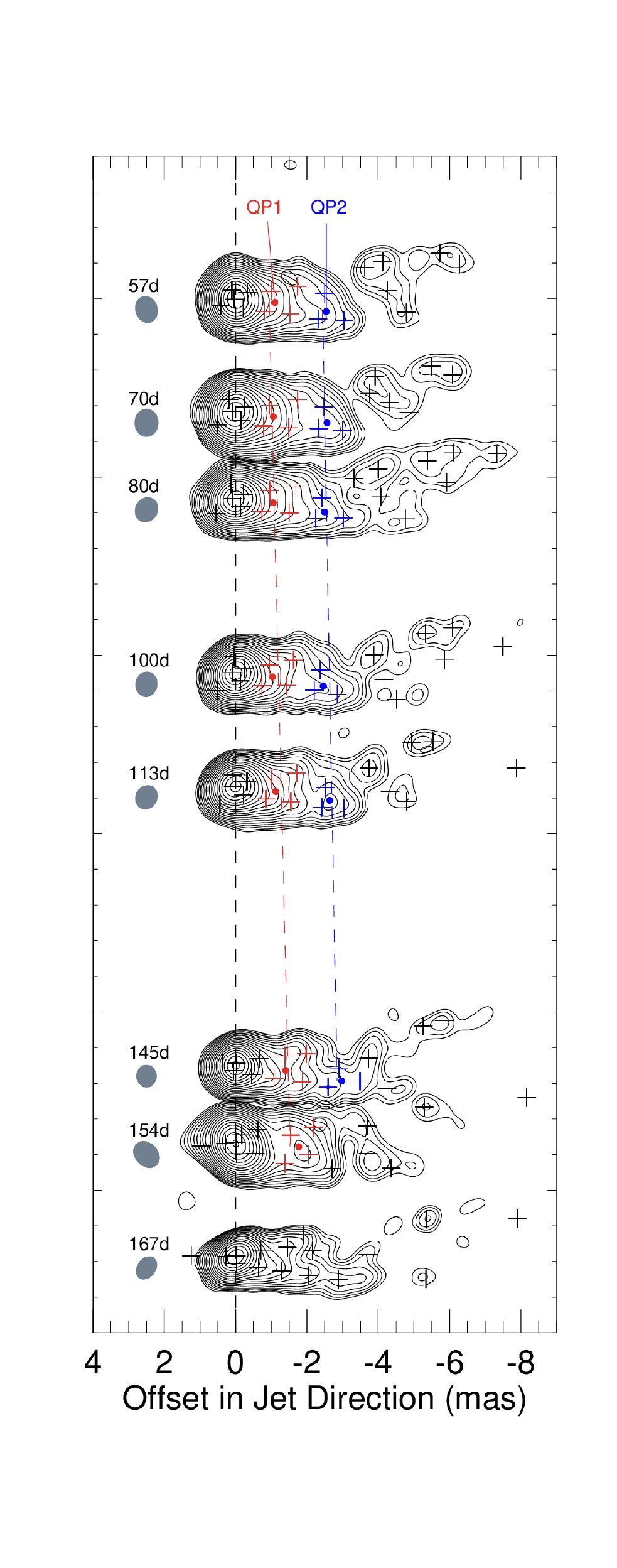}
\caption{Same as Figure~\ref{cleank} but at 43 GHz. Contours start at 2.2 mJy per beam and increase by factors of $\sqrt{2}$.
\label{cleanq}}
\end{center}
\end{figure*}

\subsection{Modelfit with Circular Gaussian Components}
\label{circularGaussian}

We fitted several circular Gaussian components to the visibility data with the task {\tt modelfit} in Difmap. We added new components if it decreases the reduced $\chi^2$ of the fit and if the components do not become point-like. With these criteria, the total number of components were about five and four for the 22 and 43 GHz data, respectively. We found that those circular Gaussian components could reproduce the overall jet structure well with the peak intensity in the residual images less than 10 mJy per beam at both observing frequencies. This is similar to previous observations by \cite{Hada2017} in 2013--2014. We present the fitted Gaussian components on top of the CLEAN maps at 22 and 43 GHz in the left panels of Figures~\ref{cleank} and~\ref{cleanq}, respectively. We cross-identified the components in different epochs based on the assumption that the flux density, size, and position of the components should not change abruptly between adjacent epochs. At 22 GHz, the distributions of different identified components, labeled as KG1, KG2, KG3, and KG4, in different epochs are quite similar. We did not identify KG1 in the last epoch because another component was detected at $\approx0.5$ mas, which made it impossible to cross-identify with the earlier epochs. We also note that we did not attempt to model the evolution of the highly complex jet structures at $\approx20$ mas using this method. We fitted a linear function to the separation from the core with time for each identified component, obtaining jet velocities at different distances. At 43 GHz, we identify three components, labeled as QG1, QG2, and QG3, showing similar distributions in different epochs. However, we did not identify the components in the last epoch due to their abrupt changes in flux densities and positions compared to earlier epochs. In addition, the component at $\approx0.5$ mas in the first epoch has a size much smaller than the later epochs, and we did not include this component.

\begin{figure}[t!]
\begin{center}
\includegraphics[trim=2mm 12mm 7mm 8mm, clip, width = 0.48\textwidth]{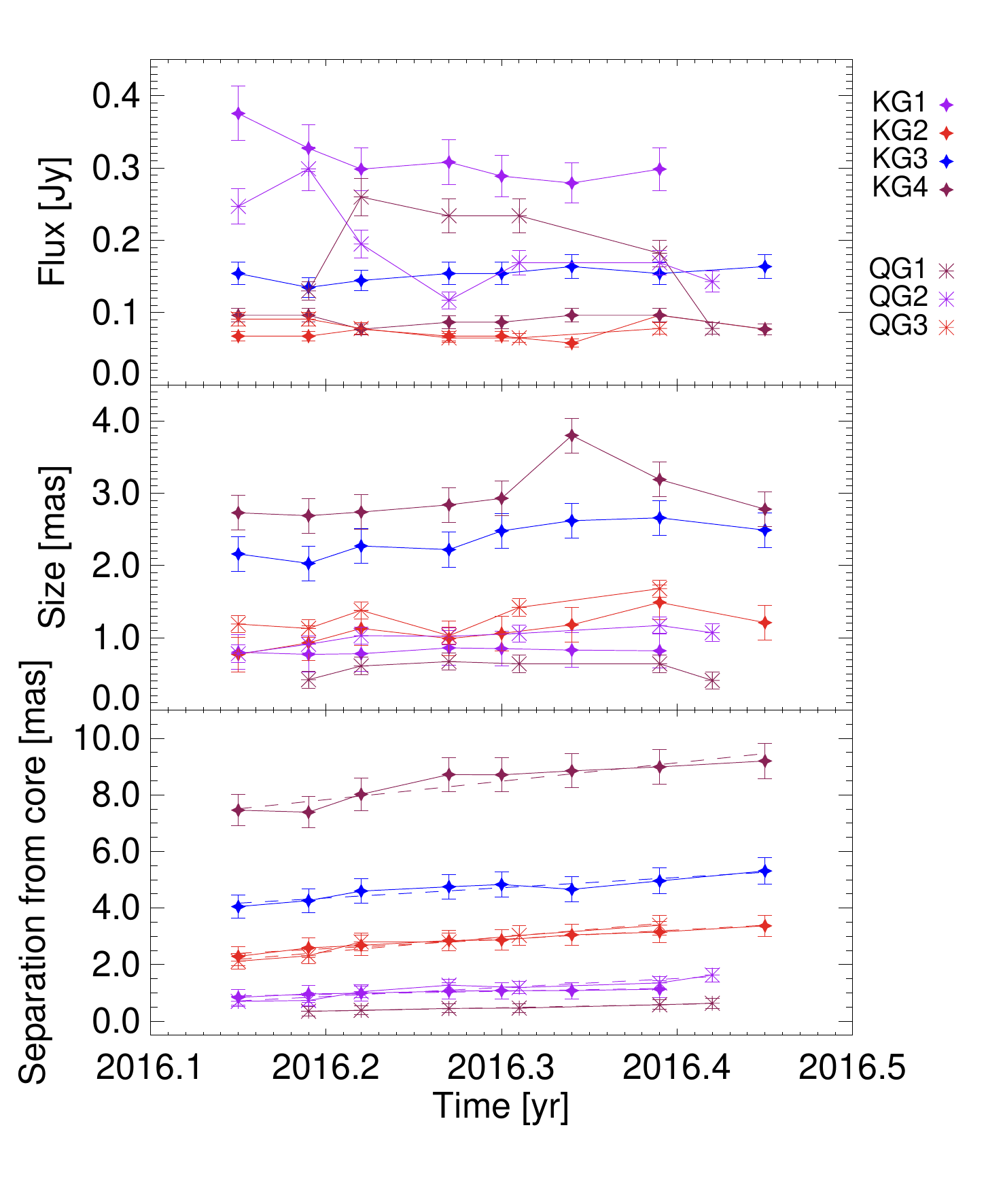}
\caption{Flux density (top), size (middle), and separation from core (bottom) as functions of time for different identified components in the {\tt modelfit} analysis with circular Gaussian components (Section~\ref{circularGaussian}) at 22 (diamonds) and 43 GHz (asterisks). The names of identified components are noted in the top right. The dashed lines in the bottom panel show the best-fit lines.
\label{gauprop}}
\end{center}
\end{figure}

We present the flux, size, and separation from the core as functions of time for the identified components in Figure~\ref{gauprop}. The properties of each identified component vary smoothly in general, suggesting that the components in different epochs may trace the same part of the jet. Remarkably, separation from the core for the components KG1 and QG2, and KG2 and QG3 are consistent with each other, gradually increasing with time.

Estimating the errors of component position is not straightforward. We assumed that the error is one-fifth of the beam size at the map center. The errors linearly increase with distance to become comparable to the beam size at the distances of the observed maximum jet extension, $\approx25$ mas and $\approx 7$ mas at 22 and 43 GHz, respectively. This approach is based on the fact that the position errors of faint components at larger jet distances would be larger than those of bright components close to the core \citep{Fomalont1999, Lee2008} and that a similar approach was adopted in previous studies of jet collimation (e.g., \citealt{Mertens2016}).

\begin{figure}[t!]
\begin{center}
\includegraphics[trim=2mm 22mm 7mm 9mm, clip, width = 0.48\textwidth]{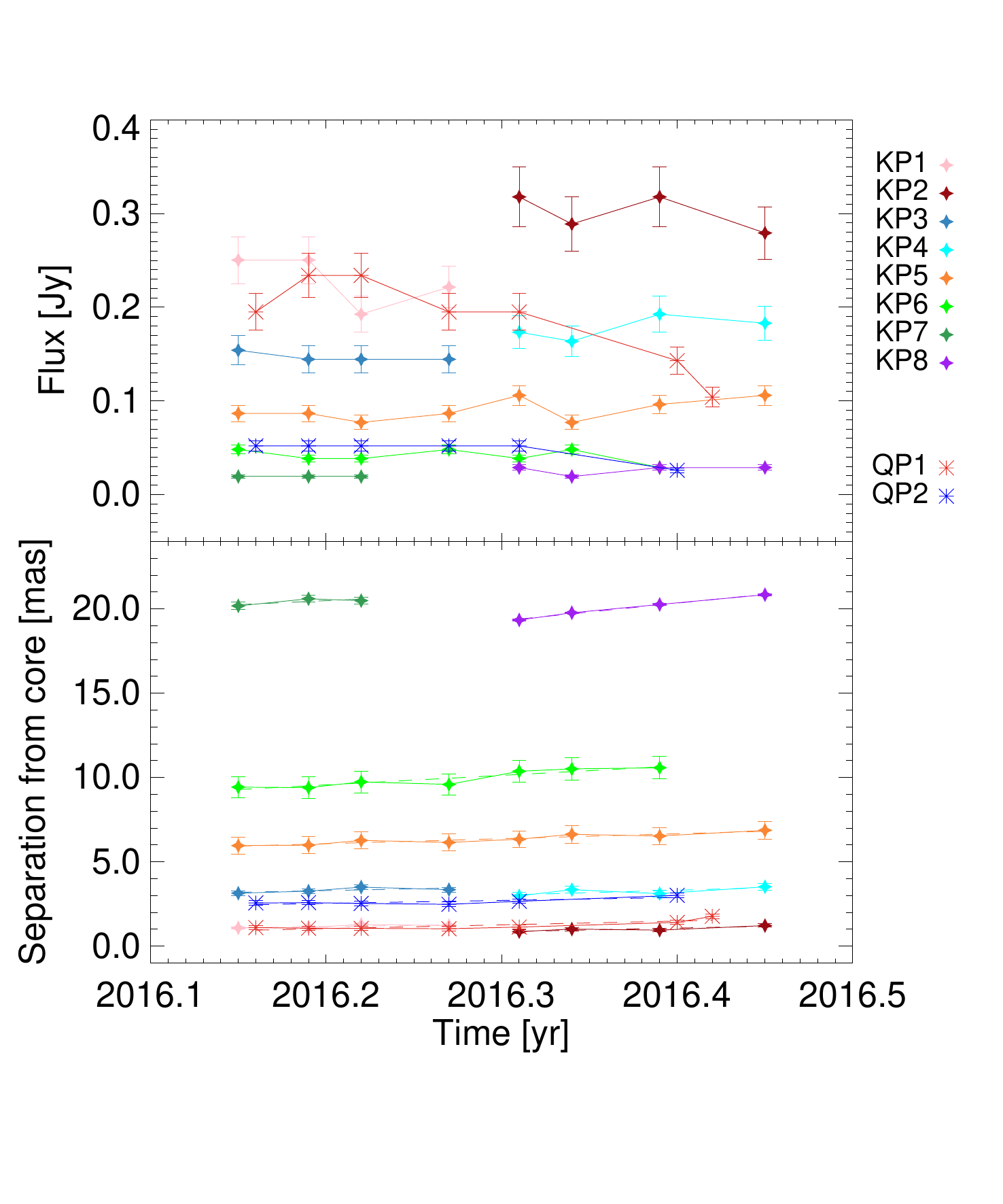}
\caption{Flux density (top) and separation from core (bottom) as functions of time for different identified components in the {\tt modelfit} analysis with point source components (Section~\ref{point}) at 22 (diamonds) and 43 GHz (asterisks). The names of identified components are noted in the top right. The dashed lines in the bottom panel show the best-fit lines.
\label{pointprop}}
\end{center}
\end{figure}

\subsection{Modelfit with Point Sources and Grouping}
\label{point}

Since the jet shows a complex structure, fitting a small number of circular Gaussian components cannot reproduce the observed jet emission accurately, which may introduce additional errors in the kinematics analysis. Thus, we increase the number of model components this time until the residual images become dominated by noise, similar to the usual {\tt modelfit} analysis that has been applied to the jet kinematics of many radio-loud AGNs (e.g., \citealt{Lister2016, Jorstad2017}). During the {\tt modelfit} procedure, we found that the sizes of many Gaussian components became point-like. The distribution of those components with zero size changes from epoch to epoch, which makes component identification difficult. Thus, we used point source components with zero sizes instead of using circular Gaussian components to be consistent in different epochs. The distribution of fitted components on top of naturally-weighted CLEAN images at 22 and 43 GHz is shown in the right panels of Figures~\ref{cleank} and~\ref{cleanq}, respectively.

We also introduced a grouping of different point source components for the jet kinematics and component identification. This is because each point source model may not represent a distinct jet emission region, which must have a finite size, although fitting with many point source models would reproduce the observed jet structures quite well mathematically. As we are interested in obtaining the jet bulk speeds, grouping different components could be a good strategy for analyzing complex jet structures (see, e.g., \citealt{Lisakov2017} for the case of 3C 273). We obtained the positions of grouped components by averaging the positions of individual components weighted by their flux densities.

We adopt different grouping schemes for different jet regions. We begin with a detailed explanation of the 22 GHz data. There are six and seven components at $\lesssim5$ mas in the first four and the last four epochs, respectively. We found that the distribution of these components is very similar in the first and the last four epochs, separately. Thus, we group two components at $\approx1$ mas and identify them as a single component in the two periods separately (KP1, KP2). The four and five components at $\approx3$ mas are grouped and identified in the two periods separately as well (KP3, KP4). This grouping scheme is based on our assumption that each grouped component represents the same jet region, provided that individual components used for the grouping have the same total number of components and a similar spatial distribution in different epochs. At a distance of $\approx8-13$ mas, a triangle-like jet shape is detected in all epochs except the last epoch, which led us to group and to identify the components in this region (KP6). Then, the remaining components at $\approx5-8$ mas are grouped and identified (KP5).

At a distance $\approx20$ mas, the jet re-brightens and significant emission is detected in all epochs. The shape of this emitting region is arc-like, which is reminiscent of the filamentary jet structures detected on pc scales (e.g., \citealt{Reid1989, Walker2018}) and on kpc scales (e.g., \citealt{Owen1989, Perlman2001, Lobanov2003}). The bright knot in the southern limb at $\approx20$ mas apparently moves outward from the first to third epochs, moves inward from the third to the fifth epochs, and then moves outward again in the last four epochs. The inward motion corresponds to an apparent speed of $\approx-3c$, which indicates that the bright knot in the first three and in the last four epochs may not be the same parts of the jet. Based on this assumption, we group and identify the components associated with the bright knot in the first three epochs and in the last four epochs separately (KP7 and KP8).

At 43 GHz, we could detect four point source components at 1--2 mas and three components at 2--3 mas in many epochs. We group and identify these components and obtain the jet speeds. We present the flux density and separation from the core as functions of time for different grouped components in Figure~\ref{pointprop}. Similarly to the case of our {\tt modelfit} analysis with circular Gaussian components (Section~\ref{circularGaussian}), the properties of the grouped components vary smoothly, indicating that they may trace the same parts of the jet in different epochs. We estimate the errors of the positions of grouped components as follows. For those that have the same numbers and similar distributions of individual components in different epochs, i.e., KP1, KP2, KP3, KP4, QP1, and QP2, we assumed the errors which provide us with $\chi^2/d.o.f. = 1$ for the fitting of linear functions to the separation from the core, where $d.o.f.$ denotes the degree of freedom. This approach was also applied to KP7 and KP8, for which we traced the components associated with the bright knot at $\approx20$ mas. This is because there would not be much errors introduced by the grouping or component identification in these cases. For other grouped components, we estimated the position errors in the same manner as in Section~\ref{circularGaussian}.

\begin{figure*}[t!]
\begin{center}
\includegraphics[trim=0mm -5mm 0mm 0mm, clip, width = 0.58\textwidth]{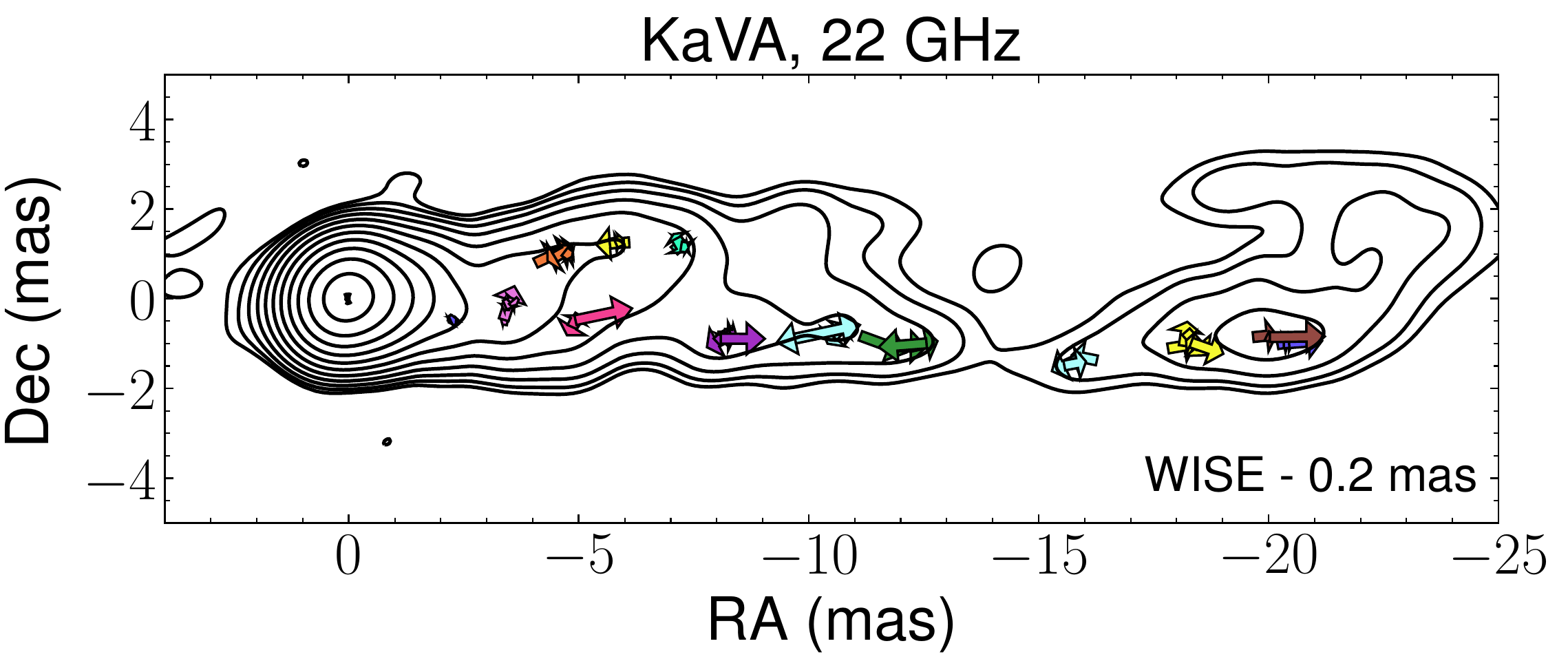}
\includegraphics[trim=0mm 0mm 0mm 0mm, clip, width = 0.415\textwidth]{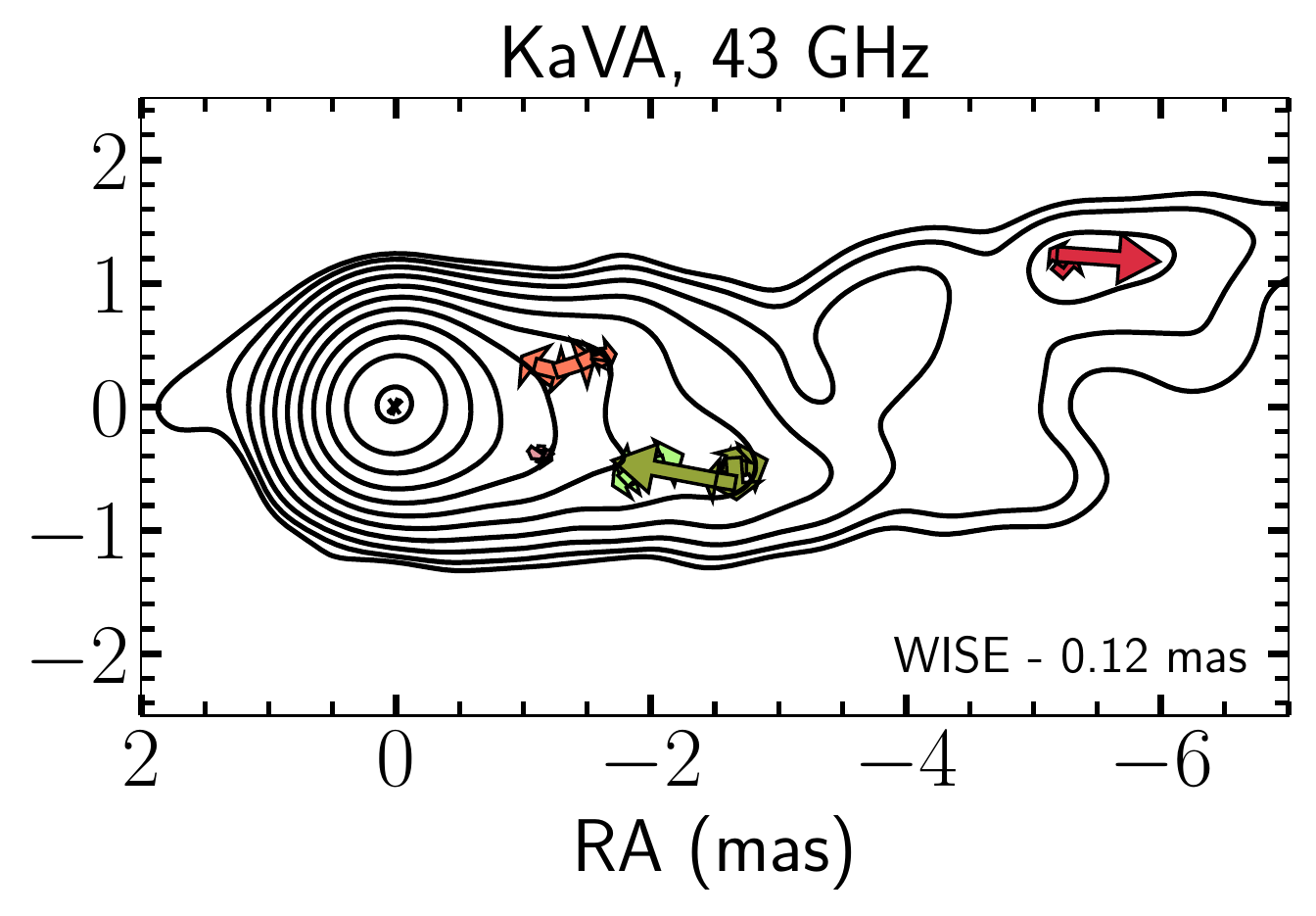}
\caption{Stacked maps of the CLEAN images at 22 (left) and 43 GHz (right) rotated clockwise by $18^\circ$ with respect to the map center. Contours start at 0.8 mJy per beam and increase by factors of $\sqrt{2}$. The displacements of several SSPs between adjacent epochs detected at the finest SWD scales are shown with the arrows (see Section~\ref{sectwise} for more details). \label{wise}}
\end{center}
\end{figure*}

\subsection{WISE}
\label{sectwise}

We applied the WISE analysis technique \citep{ML2015} to our KaVA images of the M87 jet. WISE uses 2D cross-correlations to statistically determine the jet kinematical structures without the limitations of using pre-defined templates for the brightness distributions (such as 2D Gaussian distributions) of the structures. We decomposed each map into sub-components with the segmented wavelet decomposition (SWD) method to detect a set of SSPs with a $3\sigma$ detection threshold. We implemented the intermediate wavelet decomposition (IWD\footnote{IWD allows one to cover intermediate scales between SWD scales and to improve cross identification of the individual features (see Appendix A in \citealt{Mertens2016})}) as well for a robust detectability of displacements of SSPs \citep{Mertens2016}. Determining appropriate SWD/IWD scales is important because WISE detects jet velocities by matching SSPs between adjacent epochs in such a way that the matching on smaller scales is regulated by that on larger scales (so-called multi-scale cross-correlation, MCC, \citealt{ML2015}). The maximum scale determines the largest displacement of SSPs between adjacent epochs and should be large enough to be comparable to the largest expected displacement. The minimum scale should be small enough to detect motions of fine-scale structures, while it should be larger than the minimum resolvable size of the images, which is dependent on the signal-to-noise ratio \citep{Lobanov2005}.

We assumed the minimum scale to be $\approx1/5$ of the beam size and maximum scale to be about one beam size. We applied the SWD on three spatial scales of 0.2 (0.12), 0.4 (0.24) and 0.8 (0.48) mas and amended them with the IWD on scales of 0.3 (0.18), 0.6 (0.36), and 1.2 (0.72) mas at 22 (43) GHz. We identify SSPs in adjacent epochs by using the MCC method with a tolerance factor of 1.5 and a correlation threshold of 0.7. In the MCC analysis, we search for velocity vectors with symmetric search windows of [$-31$, $+31$] mas/yr and [$-18$, $+18$] mas/yr in a longitudinal direction relative to the jet axis at 22 and 43 GHz, respectively. We used search windows of [$-10$, $+10$] mas/yr and [$-5$, $+5$] mas/yr in a transverse direction at 22 and 43 GHz, respectively. This constraint is based on the maximum SWD/IWD scales and the typical time separation between adjacent epochs. We use velocity vectors of the SSPs matched in at least four successive epochs for our further analysis. We obtained the apparent speeds by fitting linear functions to their separation from the core with time. 

We note that we used symmetric velocity search windows, while previous studies using WISE analysis adopted asymmetric search windows in a longitudinal direction \citep{Mertens2016, Boccardi2019}. The reason for using asymmetric windows in those studies is based on our prior knowledge that fast inward motions are not physically realistic. However, at least for the present study, we found that using an asymmetric velocity search window can lead to biased results towards detecting faster speeds. In Appendix~\ref{vlbawise}, we perform several tests by using one hundred randomly shuffled images of the VLBA 1.7 GHz data presented in Section~\ref{sect5}, which provide us with a number of detected WISE vectors and good statistics. We observed a systematic acceleration pattern with distance even for the shuffled data set when we use an asymmetric velocity search window. This indicates that an artificial acceleration feature can be observed in the real data as well due to the asymmetric search window. Thus, we used symmetric search windows to ensure that the observed speeds and jet acceleration result from real jet motions.

Also, we found that the statistical behaviors of the real data and the shuffled data become significantly different when using the SSPs matched in more than four or five successive epochs (Appendix~\ref{vlbawise}). This result implies that false detection due to a chance alignment may be suppressed in our KaVA WISE analysis that uses SSPs matched in more than four successive epochs. However, a number of jet motions are still detected in the random data set when a length-chain of four or five successive epochs is used. This suggests that one needs to use a longer length-chain such as seven or eight successive epochs, if possible, to obtain robust results. Nevertheless, we had only eight epochs in total for the KaVA images and using such a long length-chain was not available. Instead, we used a larger correlation threshold (0.7) compared to the previous studies (0.6, \citealt{Mertens2016, Boccardi2019}) for our KaVA data to reduce the probability of false detection.



\begin{figure*}[t!]
\begin{center}
\includegraphics[trim=4mm 4mm 7mm 9mm, clip, width = 0.48\textwidth]{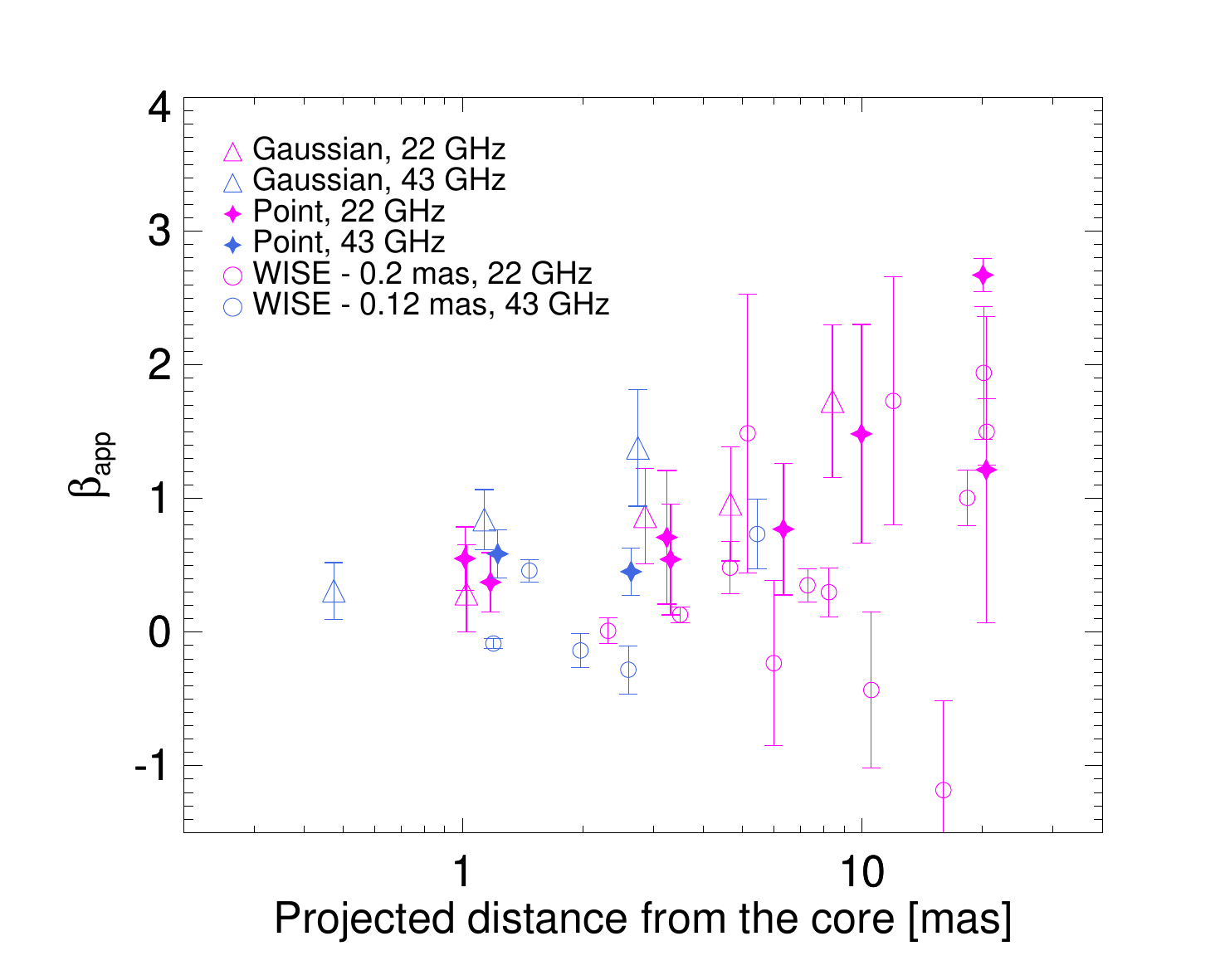}
\includegraphics[trim=4mm 4mm 7mm 9mm, clip, width = 0.48\textwidth]{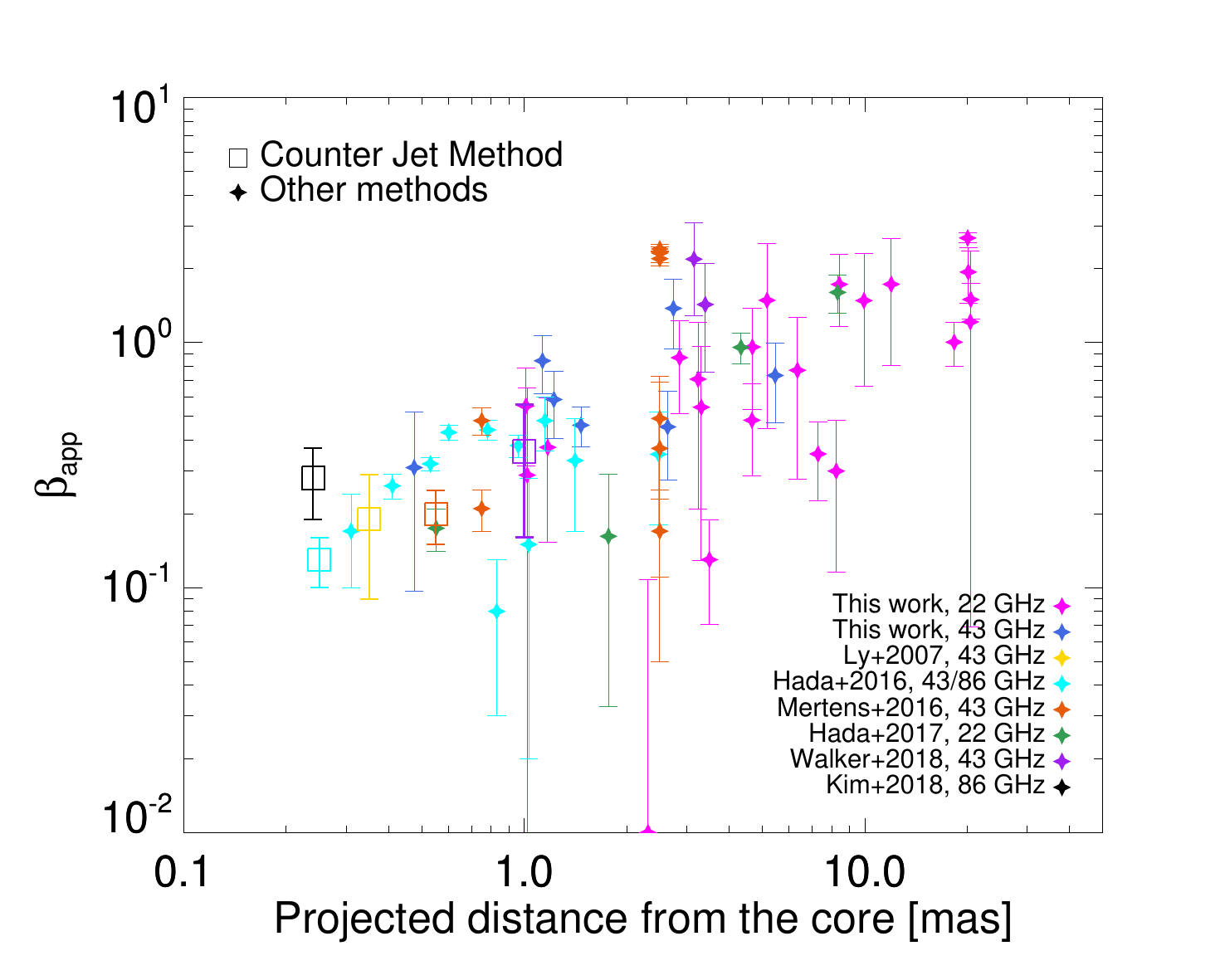}
\caption{Left: apparent speed as a function of projected distance from the core from the KaVA observations at 22 (magenta) and 43 (blue) GHz. The different symbols are from different methods applied to the jet kinematics (Section~\ref{sect4}). Right: same as the left panel but adding the apparent speeds obtained from other studies to the diagram \citep{Ly2007, Hada2016, Mertens2016, Hada2017, Walker2018, Kim2018}. The data points obtained by the jet to counterjet brightness ratio analysis are shown in the open squares, while those obtained by other methods are shown in the filled diamonds. We note that the negative speeds detected by the WISE analysis (but consistent with zero speeds within $2\sigma$) shown in the left panel are not drawn in this logarithmic diagram.
\label{vel}}
\end{center}
\end{figure*}

We found that our WISE results show a significant dependence on different SWD/IWD scales at both frequencies (see Appendix~\ref{kavawise} for more details). This behavior was not clearly observed in the previous WISE analysis of the VLBA data at 43 GHz \citep{Mertens2016}, nor in our WISE analysis of the VLBA data at 1.7 GHz (Section~\ref{sect5}). The scale dependence is likely due to the limited angular resolution of our KaVA observations because the sampling interval of our data ($\approx2$ weeks) is short enough compared to the previous studies ($\gtrsim3$ weeks, \citealt{Mertens2016, Boccardi2019}). Based on this argument, we selected the results of the finest scales, which provides the highest effective spatial resolution among different scales. Figure~\ref{wise} shows the WISE vectors detected at the finest scales. We found many vectors having a wide range of speed from stationary motions to fast outward motions up to $\approx2c$.

\subsection{Jet Apparent Speeds and Comparison with Other Studies \label{sect44}}

We show the mean radial distances from the core and the observed radial speeds of the jet components obtained by the three methods in Table~\ref{result}. We present the apparent jet speeds in the left panel of Figure~\ref{vel}. The apparent speeds, in general, increase from $\approx0.3c$ at a distance $\approx0.5$ mas to $\approx2.7c$ at $\approx20$ mas. However, there is non-negligible dispersion in the speeds at a given distance. Many of the slower speeds come from the WISE analysis. On the contrary, the speeds of the fast outward WISE vectors are in good agreement with the {\tt modelfit} results. This may imply that the WISE analysis can capture both fast and slow motions existing in the M87 jet, as already seen by \cite{Mertens2016}, which is not possible for our {\tt modelfit} results because we measured a single velocity for each jet region.

However, we cannot reject the possibility that the dispersion is produced by potential systematic errors in the different methods that we have applied. As previously explained in Section~\ref{sect3}, the first method, {\tt modelfit} with circular Gaussian components, has the advantage of a straightforward component identification in individual epochs. However, the results can be affected by non-negligible residual emission in the maps which could not be properly modeled by only a few Gaussian components. The second method, {\tt modelfit} with point sources and grouping of different components for cross-identification in different epochs, is not affected by the residual jet emission. However, it relies on visual inspection for grouping and identification and could be quite subjective. The WISE analysis is, on the other hand, based on statistical methods. However, we could not figure out the origin of the dependence on SWD/IWD scales and we selected the results at the finest scales.

\begin{deluxetable}{cccc}[t!]

\tablecaption{Results of jet kinematics}
\tablehead{
\colhead{Methods} & \colhead{$\langle R \rangle$ (mas)} &
\colhead{$\mu_r$ (mas $\rm yr^{-1}$)} & \colhead{$\beta_{\rm app}$}  \\
(1) & (2) & (3) & (4)
}
\startdata
\multicolumn{4}{c}{KaVA 22 GHz} \\
\midrule
\multirow{4}{0.25\columnwidth}{\centering Circular Gaussian}  & $1.02\pm0.10$ & $1.09\pm1.38$ & $0.29\pm0.37$ \\
 &  $2.86\pm0.34$ &  $3.28\pm1.35$ &  $0.87\pm0.36$ \\
 &  $4.68\pm0.39$ &  $3.63\pm1.61$ & $0.96\pm0.43$ \\
 &  $8.42\pm0.70$ &  $6.54\pm2.15$ &  $1.73\pm0.57$ \\
  \midrule
 \multirow{8}{0.25\columnwidth}{\centering Point Sources}  & 
      $1.01\pm0.16$ &  $2.08\pm0.90$ &  $0.55\pm0.24$ \\
 &  $1.17\pm0.09$ & $1.41\pm0.84$ & $0.37\pm0.22$ \\
 &  $3.31\pm0.15$ &  $2.06\pm1.57$ & $0.54\pm0.42$ \\
 &  $3.24\pm0.24$ &  $2.68\pm1.89$ &  $0.71\pm0.50$ \\
 &  $6.34\pm0.32$ & $2.92\pm1.87$ & $0.77\pm0.49$ \\
 &  $9.95\pm0.52$ &  $5.61\pm3.10$ &  $1.48\pm0.82$ \\
 &  $20.05\pm0.64$ &  $10.05\pm0.47$ & $2.67\pm0.13$ \\
 &  $20.42\pm0.22$ &  $4.57\pm4.31$ &  $1.21\pm1.15$ \\
 \midrule
 \multirow{13}{0.25\columnwidth}{\centering WISE}  & 
     $2.31\pm0.02$ & $0.04\pm0.37$ & $0.01\pm0.10$ \\
 &  $3.50\pm0.07$ & $0.49\pm0.22$ & $0.13\pm0.06$ \\
 &  $4.66\pm0.25$ & $1.81\pm0.74$ & $0.48\pm0.20$ \\
 &  $5.16\pm0.36$ & $5.60\pm3.92$ & $1.49\pm1.04$ \\
 &  $6.00\pm0.15$ & $-0.87\pm2.32$ & $-0.23\pm0.62$ \\
 &  $7.29\pm0.09$ & $1.32\pm0.47$ & $0.35\pm0.12$ \\
 &  $8.25\pm0.20$ & $1.12\pm0.69$ & $0.30\pm0.18$ \\
 &  $10.53\pm0.36$ & $-1.63\pm2.20$ & $-0.43\pm0.59$ \\
 &  $11.96\pm0.49$ & $6.51\pm3.49$ & $1.73\pm0.93$ \\
 &  $15.96\pm0.25$ & $-4.45\pm2.51$ & $-1.18\pm0.67$ \\
 &  $18.31\pm0.28$ & $3.78\pm0.77$ & $1.00\pm0.21$ \\
 &  $20.14\pm0.43$ & $7.30\pm1.87$ & $1.94\pm0.50$ \\
 &  $20.48\pm0.25$ & $5.64\pm0.94$ & $1.50\pm0.25$ \\
 \midrule
 \multicolumn{4}{c}{KaVA 43 GHz} \\
\midrule
\multirow{3}{0.25\columnwidth}{\centering Circular Gaussian}  & $0.47\pm0.11$ & $1.17\pm0.80$ & $0.31\pm0.21$ \\
 &  $1.13\pm0.34$ &  $3.19\pm0.85$ &  $0.84\pm0.23$ \\
 &  $2.74\pm0.46$ &  $5.22\pm1.65$ & $1.38\pm0.44$ \\
 \midrule
\multirow{2}{0.25\columnwidth}{\centering Point Sources}  & 
     $1.22\pm0.27$ &  $2.22\pm0.68$ &  $0.59\pm0.18$ \\
&  $2.64\pm0.19$ & $1.71\pm0.67$ & $0.45\pm0.18$ \\
 \midrule
 \multirow{5}{0.25\columnwidth}{\centering WISE}  & 
     $1.19\pm0.04$ & $-0.32\pm0.15$ & $-0.09\pm0.04$ \\
 &  $1.47\pm0.20$ & $1.73\pm0.32$ & $0.46\pm0.08$ \\
 &  $1.97\pm0.09$ & $-0.52\pm0.47$ & $-0.14\pm0.13$ \\
 &  $2.60\pm0.21$ & $-1.06\pm0.68$ & $-0.28\pm0.18$ \\
 &  $5.46\pm0.20$ & $2.76\pm0.99$ & $0.73\pm0.26$ \\
\midrule
 \multicolumn{4}{c}{VLBA 1.7 GHz} \\
\midrule
\multirow{8}{0.25\columnwidth}{\centering Circular Gaussian}  & $344.07\pm2.43$ & $7.30\pm7.36$ & $1.93\pm1.94$ \\
 &  $353.07\pm5.02$ &  $9.13\pm3.04$ &  $2.41\pm0.80$ \\
 &  $368.32\pm7.23$ &  $12.88\pm3.11$ &  $3.40\pm0.82$ \\
 &  $370.73\pm3.83$ &  $7.96\pm4.05$ &  $2.10\pm1.07$ \\
 &  $378.98\pm7.00$ &  $12.65\pm3.17$ &  $3.34\pm0.84$ \\
 &  $381.14\pm8.58$ &  $15.91\pm2.86$ &  $4.20\pm0.76$ \\
 &  $404.47\pm10.68$ &  $19.22\pm3.30$ &  $5.08\pm0.87$ \\
 &  $406.11\pm3.37$ &  $11.92\pm9.13$ &  $3.15\pm2.41$\\
\midrule
\multirow{14}{0.25\columnwidth}{\centering WISE}  & 
     $   9.51 \pm  0.77 $ & $  0.89 \pm  0.22 $ & $  0.24 \pm  0.06 $ \\
 & $  11.58 \pm  0.85 $ & $  1.11 \pm  1.01 $ & $  0.30 \pm  0.27 $ \\
 & $  18.63 \pm  1.12 $ & $  0.21 \pm  0.22 $ & $  0.06 \pm  0.06 $ \\
 & $  29.35 \pm  4.24 $ & $  6.22 \pm  1.50 $ & $  1.65 \pm  0.40 $ \\
 & $  32.01 \pm  2.05 $ & $  1.78 \pm  0.57 $ & $  0.47 \pm  0.15 $ \\
 & $  43.00 \pm  0.66 $ & $ -0.14 \pm  0.87 $ & $ -0.04 \pm  0.23 $ \\
 & $  50.13 \pm  3.73 $ & $  4.37 \pm  2.67 $ & $  1.16 \pm  0.71 $ \\
 & $  52.49 \pm  2.25 $ & $  5.89 \pm  1.38 $ & $  1.57 \pm  0.37 $ \\
 & $  63.52 \pm  1.29 $ & $  0.86 \pm  0.46 $ & $  0.23 \pm  0.12 $ \\
 & $  66.07 \pm  2.15 $ & $ -0.67 \pm  2.83 $ & $ -0.18 \pm  0.75 $ \\
 & $  74.91 \pm  4.42 $ & $ -0.33 \pm  1.05 $ & $ -0.09 \pm  0.28 $ \\
 & $  83.58 \pm  4.41 $ & $ -6.66 \pm  1.29 $ & $ -1.77 \pm  0.34 $ \\
 & $  89.32 \pm  6.11 $ & $ -2.03 \pm  1.35 $ & $ -0.54 \pm  0.36 $ \\
     &  $\vdots$ & $\vdots$ & $\vdots$
\enddata
\tablecomments{(1) Methods of the jet kinematics used. (2) Mean radial distance from the core and $1\sigma$ scatter of the distances. (3) Angular radial speed and $1\sigma$ uncertainty. (4) Radial speed in units of the speed of light and $1\sigma$ uncertainty. \label{result}}
\end{deluxetable}

\addtocounter{table}{-1}

\begin{deluxetable}{cccc}[t!]

\tablecaption{Continued}
\tablehead{
\colhead{Methods} & \colhead{$\langle R \rangle$ (mas)} &
\colhead{$\mu_r$ (mas $\rm yr^{-1}$)} & \colhead{$\beta_{\rm app}$}  \\
(1) & (2) & (3) & (4)
}
\startdata
 \multicolumn{4}{c}{VLBA 1.7 GHz} \\
\midrule
\multirow{37}{0.25\columnwidth}{\centering WISE}  &
 $\vdots$ & $\vdots$ & $\vdots$\\
 & $  97.71 \pm  0.60 $ & $  0.14 \pm  0.39 $ & $  0.04 \pm  0.10 $ \\
 & $ 110.68 \pm  0.48 $ & $  1.10 \pm  0.39 $ & $  0.29 \pm  0.10 $ \\
 & $ 112.22 \pm  4.99 $ & $ -6.04 \pm  3.51 $ & $ -1.61 \pm  0.93 $ \\
 & $ 128.24 \pm  1.71 $ & $ -3.30 \pm  1.72 $ & $ -0.88 \pm  0.46 $ \\
 & $ 136.05 \pm  7.76 $ & $  7.22 \pm  2.24 $ & $  1.92 \pm  0.60 $ \\
 & $ 138.29 \pm  1.42 $ & $ -0.76 \pm  0.82 $ & $ -0.20 \pm  0.22 $ \\
 & $ 154.85 \pm  6.22 $ & $  8.14 \pm  0.53 $ & $  2.17 \pm  0.14 $ \\
 & $ 170.15 \pm  6.07 $ & $  7.33 \pm  0.69 $ & $  1.95 \pm  0.18 $ \\
 & $ 180.72 \pm  6.74 $ & $  7.55 \pm  0.72 $ & $  2.01 \pm  0.19 $ \\
 & $ 189.97 \pm  1.61 $ & $  1.17 \pm  0.70 $ & $  0.31 \pm  0.19 $ \\
 & $ 220.33 \pm  2.71 $ & $ -1.78 \pm  2.27 $ & $ -0.47 \pm  0.60 $ \\
 & $ 229.16 \pm  2.24 $ & $  1.42 \pm  1.88 $ & $  0.38 \pm  0.50 $ \\
 & $ 230.84 \pm  6.52 $ & $  7.03 \pm  1.41 $ & $  1.87 \pm  0.37 $ \\
 & $ 234.93 \pm 13.52 $ & $ 11.61 \pm  1.15 $ & $  3.09 \pm  0.31 $ \\
 & $ 251.49 \pm  3.16 $ & $  2.83 \pm  2.14 $ & $  0.75 \pm  0.57 $ \\
 & $ 254.01 \pm  0.98 $ & $  2.18 \pm  0.87 $ & $  0.58 \pm  0.23 $ \\
 & $ 256.03 \pm  2.56 $ & $  0.29 \pm  1.05 $ & $  0.08 \pm  0.28 $ \\
 & $ 256.77 \pm  5.24 $ & $  9.63 \pm  0.44 $ & $  2.56 \pm  0.12 $ \\
 & $ 265.96 \pm  3.88 $ & $  4.66 \pm  1.05 $ & $  1.24 \pm  0.28 $ \\
 & $ 266.95 \pm  2.61 $ & $ -1.18 \pm  1.17 $ & $ -0.32 \pm  0.31 $ \\
 & $ 270.88 \pm  3.37 $ & $ -7.25 \pm  1.62 $ & $ -1.93 \pm  0.43 $ \\
 & $ 279.45 \pm  5.06 $ & $ 10.55 \pm  1.23 $ & $  2.81 \pm  0.33 $ \\
 & $ 288.37 \pm  1.80 $ & $  2.80 \pm  1.48 $ & $  0.74 \pm  0.39 $ \\
 & $ 288.81 \pm  4.83 $ & $  9.60 \pm  1.79 $ & $  2.55 \pm  0.47 $ \\
 & $ 292.54 \pm  8.65 $ & $ 16.64 \pm  1.86 $ & $  4.42 \pm  0.50 $ \\
 & $ 301.23 \pm 12.49 $ & $ 15.46 \pm  1.62 $ & $  4.11 \pm  0.43 $ \\
 & $ 309.00 \pm  5.99 $ & $ 17.43 \pm  1.41 $ & $  4.63 \pm  0.37 $ \\
 & $ 312.54 \pm  5.92 $ & $ 10.75 \pm  0.85 $ & $  2.86 \pm  0.23 $ \\
 & $ 321.37 \pm 10.98 $ & $ 14.77 \pm  1.95 $ & $  3.93 \pm  0.52 $ \\
 & $ 323.20 \pm  3.28 $ & $  1.10 \pm  3.40 $ & $  0.29 \pm  0.90 $ \\
 & $ 335.13 \pm  3.00 $ & $ -2.22 \pm  2.72 $ & $ -0.59 \pm  0.72 $ \\
 & $ 340.97 \pm  6.98 $ & $ 12.19 \pm  2.17 $ & $  3.24 \pm  0.58 $ \\
 & $ 347.63 \pm  4.14 $ & $  5.76 \pm  1.12 $ & $  1.53 \pm  0.30 $ \\
 & $ 348.68 \pm  2.42 $ & $ -2.12 \pm  1.96 $ & $ -0.56 \pm  0.52 $ \\
 & $ 375.26 \pm  5.29 $ & $ 10.44 \pm  0.76 $ & $  2.77 \pm  0.20 $ \\
 & $ 376.08 \pm  9.63 $ & $ 18.11 \pm  2.16 $ & $  4.82 \pm  0.57 $ \\
 & $ 382.17 \pm 12.44 $ & $ 21.86 \pm  1.48 $ & $  5.81 \pm  0.39 $
\enddata
\tablecomments{Continued.}
\end{deluxetable}


In the right panel of Figure~\ref{vel}, we compare our results with other previous VLBI observations. We include the results obtained from the data observed with relatively small sampling intervals (e.g., $\lesssim3$ weeks) in more than five epochs to avoid possible effects of a large sampling interval on the results, or from the analysis of jet to counterjet brightness ratio. We converted the observed brightness ratio in different studies into the apparent speeds with the adopted viewing angle of $17^\circ$ and the spectral index of $\alpha=-0.7$ for the inner jet (at 0.2--1.2 mas) obtained at 22--86 GHz \citep{Hada2016}. As for the results of \cite{Mertens2016}, we included the values derived by their stacked cross-correlation analysis which are representatives of a large number of jet speeds derived by the WISE analysis. Also, we present an average of the component speeds measured at core distances greater than 1.8 mas as reported by \cite{Walker2018}.

Although there is a general scatter as noted above, all of the different studies show a consistent trend of jet acceleration. The dispersion in the observed speeds at a given distance is also present in other studies. It is notable that the observed speeds derived by the jet to counterjet brightness ratio, which are not affected by the complications of component identification, large sampling intervals, etc., are consistent with those derived by the other methods. This result suggests that the jet is moving at subluminal speeds at distances $\lesssim1$ mas from the core. We note that our results are consistent with those of \cite{Mertens2016} and \cite{Walker2018} at distances $\lesssim1$ mas, while the speeds we obtained are lower than the fastest motions of \cite{Mertens2016} and are marginally consistent with \cite{Walker2018} at distances $\gtrsim2$ mas.

\section{Jet Kinematics on Scales of $\approx200-410$ mas Based on VLBA Archival Data}
\label{sect5}

While there is rich information about the jet velocity measurements available at relatively small distances of $\lesssim20$ mas thanks to many recent studies with VLBI observations at relatively high frequencies of $\gtrsim15$ GHz, the velocity measurements at outer jet distances is still limited. \cite{Asada2014} was the first to connect the velocity fields between mas scales and arcsecond scales, indicating substantial jet acceleration from subluminal to superluminal speeds occurring at $\approx180-450$ mas. However, recent studies including our present study have found that the jet shows superluminal motions already at $\lesssim20$ mas (see Figure~\ref{vel}). Therefore, the scale where bulk jet acceleration occurs in M87 is still under debate.

\begin{figure*}[t!]
\begin{center}
\includegraphics[trim=0mm 0mm 0mm 0mm, clip, width = \textwidth]{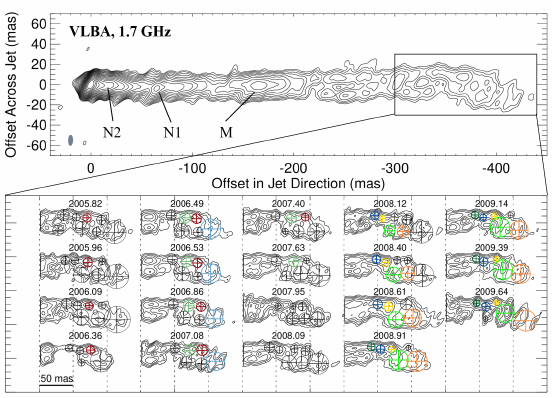}
\caption{Top: a CLEAN image of the M87 jet observed with the VLBA on 2005 Oct 27 at 1.7 GHz. The gray shaded ellipse in the lower left part denotes a typical size of the full width at half maximum of the synthesized beam. Contours start at 1 mJy per beam. The map is rotated clockwise by $23^\circ$. The re-brightened jet regions at $\approx20$, $\approx65$, and $\approx165$ mas, originally labeld as components N2, N1, and M, respectively \citep{Reid1989}, are marked. Bottom: CLEAN maps showing the region in the rectangular box in the top panel for 19 VLBA archival data we analyzed (Section~\ref{sect5}). The observation date in units of decimal years is noted for each map. The circular Gaussian components fitted to the visibility data are shown with the crosses surrounded by circles. The components with the same color in different epochs are cross-identified as the same component.
\label{vlbaclean}}
\end{center}
\end{figure*}

Possible explanations for the rapid jet acceleration observed at $\gtrsim180$ mas by \cite{Asada2014} include faster jet motions that could not be traced due to the limited angular resolution (with the FWHM of the synthesized beam of $19.9\times14.6$ mas) or the large time interval between observations (about one year). To investigate this possibility and to probe the jet velocity field over a wide distance range, we performed a complementary jet kinematic analysis by using the archival VLBA monitoring data observed in 19 epochs between 2005 and 2009 at 1.7 GHz. These data were presented in previous studies of HST-1 \citep{Cheung2007, Giroletti2012, Casadio2013}. We revisited these data in our recent study of Faraday rotation in the M87 jet \citep{Park2019}, where the details of the data reduction process are shown. In this paper, we present our kinematic analysis using these data at distances $\lesssim450$ mas.

We show a naturally-weighted CLEAN image of the first epoch data in the upper panel of Figure~\ref{vlbaclean}. A typical size of the FWHM of the synthesized beam is $11\times5$ mas with a position angle of $-2^\circ$, improved by a factor of two compared to the observations of \cite{Asada2014}. A typical rms noise level is $\approx0.2$ mJy per beam. As performed for our KaVA analysis, we perform the jet kinematics with different methods.

\subsection{Modelfit with Circular Gaussian Components}

Although the entire jet structure between the core and the extended jet down to $\approx450$ mas was successfully imaged, we found that there are locally brightened jet regions at $\approx20$, 65, and 165 mas in all epochs. The positions of these regions are almost the same in different epochs, making them appear stationary, as already reported in previous studies \citep{Reid1989, Dodson2006, Asada2014}. Our VLBA data could not resolve the north and south jet limbs at $\lesssim200$ mas. In addition, we found that the jet structure between $\approx200$ and $\approx320$ mas is quite smooth. The jet re-brightening at $\approx20$, 65, and 165 mas, the limited angular resolution, and the smooth jet structure prevented us from performing the jet kinematics based on {\tt modelfit} analysis at $\lesssim320$ mas. More specifically, we found that the total number of components in adjacent epochs at similar distances is usually not preserved in this region. The distributions of neighboring {\tt modelfit} components were not similar either, possibly due to the smooth jet structure, and we could not attempt grouping of the components.

However, we found that the jet emission at $\gtrsim320$ mas could be modeled well with several circular Gaussian components along the north and south limbs separately. This is presumably thanks to the relatively large jet width in this region (e.g., \citealt{AN2012}) and the distinct jet shape like a `head' consisting of several knotty structures. We found that the fitted components maintain finite sizes and the distributions of neighboring components are frequently preserved in many successive epochs, which does not require us to perform the grouping of different components for cross-identification.

\begin{figure}[t!]
\begin{center}
\includegraphics[trim=7mm 12mm 12mm 5mm, clip, width = 0.48\textwidth]{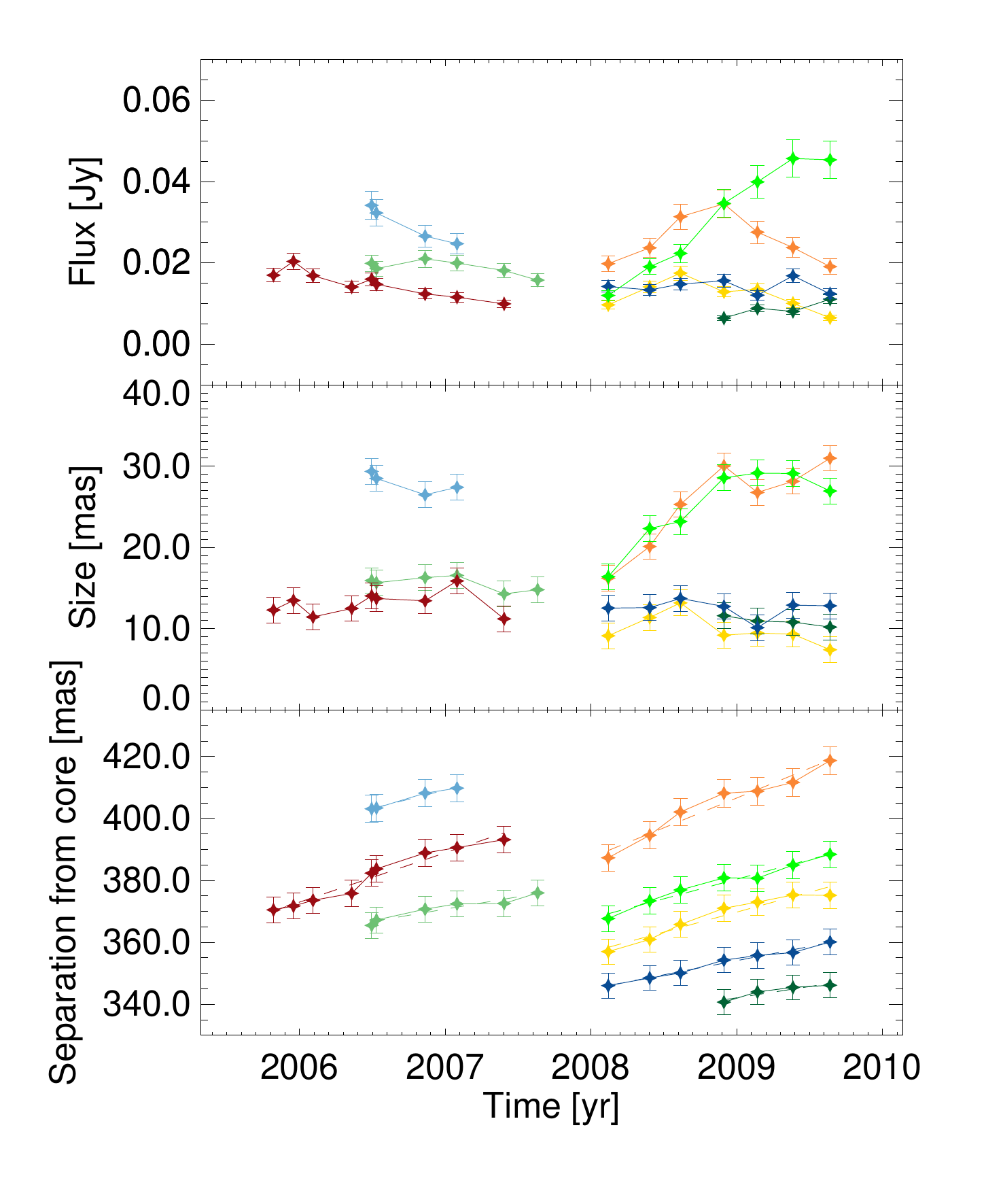}
\caption{Same as Figure~\ref{gauprop} but for the VLBA archival data observed at 1.7 GHz (Figure~\ref{vlbaclean}). The same color scheme for the identified components as in Figure~\ref{vlbaclean} is used.
\label{vlbaprop}}
\end{center}
\end{figure}

We identify components in different epochs only when (i) the distribution of neighboring components is similar and (ii) the properties of components, i.e., flux density, size, and separation from the core, vary smoothly over more than four successive epochs. The components identified to be the same part of the jet are shown with the same color in different epochs (Figure~\ref{vlbaclean}). We present the properties of these components as functions of time in Figure~\ref{vlbaprop}, showing that all the quantities vary smoothly over time. Similarly to the case of our KaVA analysis presented in Section~\ref{circularGaussian}, we assumed position errors linearly increasing from one-fifth of the synthesized beam size at a zero distance to one beam size at $\approx450$ mas. 


\begin{figure*}[t!]
\begin{center}
\includegraphics[trim=0mm 0mm 0mm 0mm, clip, width = \textwidth]{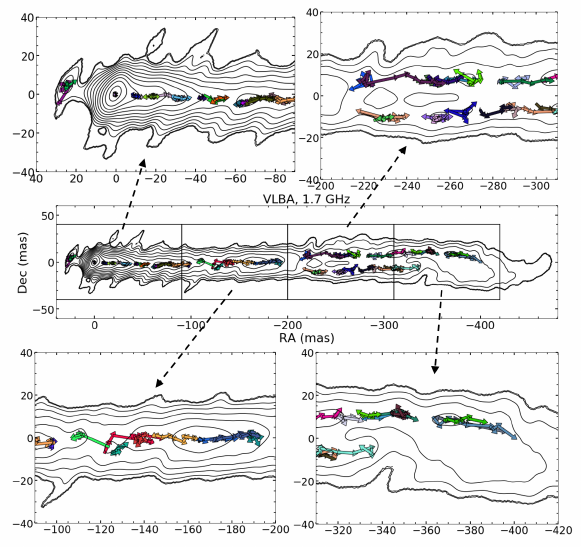}
\caption{Velocity vectors identified in more than seven successive epochs in our WISE analysis of the VLBA data on top of a stacked CLEAN image. The map is rotated clockwise by $23^\circ$. The top and bottom figures are enlarged versions of the four different jet regions shown by the rectangles in the central figure. We note that our data could not resolve the north and south jet edges at $\lesssim200$ mas, which results in most of the detected velocity vectors distributed along the central, single jet ridge.
\label{wisel}}
\end{center}
\end{figure*}

\begin{figure*}[t!]
\begin{center}
\includegraphics[trim=17mm 15mm 25mm 15mm, clip, width = 0.65\textwidth]{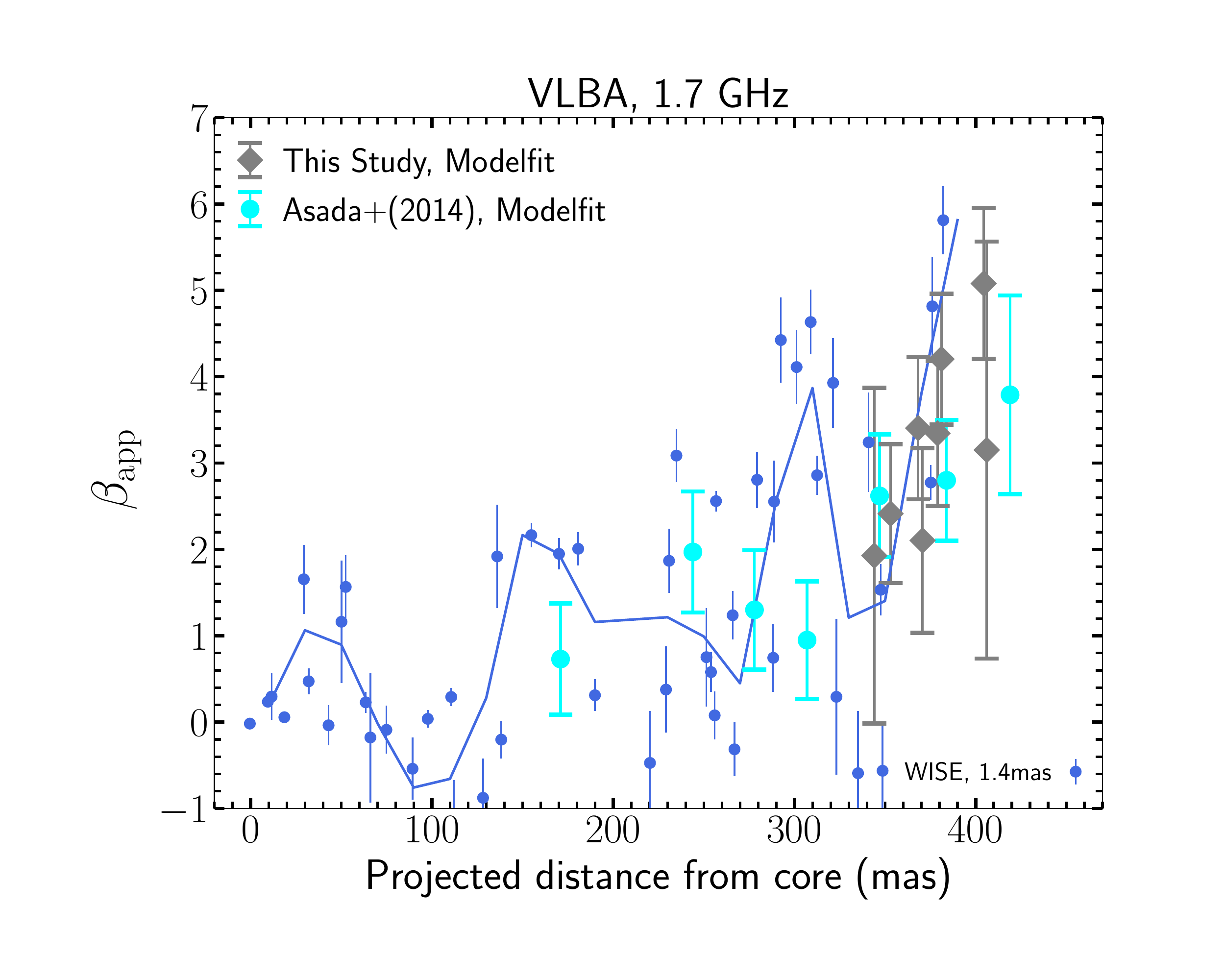}
\caption{Apparent speed as a function of projected distance from the core. The data points obtained by the WISE analysis of the VLBA data at 1.7 GHz at the finest scale are shown with the blue circles, while those obtained by the {\tt modelfit} analysis are presented with the grey filled diamonds. The blue solid line shows the data points after un-weighted binning of the WISE data points with a bin size of 20 mas. The data points taken from \cite{Asada2014} are shown with the cyan circles.
\label{wiselvel}}
\end{center}
\end{figure*}

\subsection{WISE}
\label{sectvlbawise}

We performed a WISE analysis for the VLBA 1.7 GHz data in the same manner as done for the KaVA data (Section~\ref{sectwise}). We applied the SWD on three spatial scales of 1.4, 2.8 and 5.6 mas and amended them with the IWD on scales of 2.1, 4.2, and 8.4 mas. In the MCC analysis of the VLBA data, we used a correlation threshold of 0.6, following the previous studies using WISE \citep{Mertens2016, Boccardi2019}. We constrained the ranges of speeds for matching SSPs to be between $-58$ and +58 mas/year in a longitudinal direction relative to the jet axis and between $-15$ and +15 mas/yr in a transverse direction. We use velocity vectors of the SSPs matched in at least seven successive epochs for our further analysis (see Section~\ref{sectwise} and Appendix~\ref{vlbawise} for related discussion of velocity search windows and of length-chains).


In Figure~\ref{wisel}, we present the displacement vectors detected at the finest scale of 1.4 mas. We note that we could not perform a direct one-to-one comparison between the vectors detected on different SWD/IWD scales due to their large numbers (e.g., 53 vectors identified on 1.4 mas scale). However, their general trends of apparent speed as functions of distance are in good agreement with each other (see Appendix~\ref{vlbawise}). Thus, we select the results of the finest scale as our representative WISE result for the VLBA data and use them for our further analysis.

The displacement vectors are distributed over the whole distance range where the jet emission is significant. As noted above, the north and south jet limbs could not be resolved by our data at $\lesssim200$ mas and most of the WISE vectors in this region are distributed along the central ridge. On the contrary, the vectors along the two limbs are separately detected at $\gtrsim200$ mas. WISE successfully captures jet motions at distances $\lesssim320$ mas, where it was difficult to cross-identify each part of the jet in adjacent epochs for the {\tt modelfit} analysis. It is clear that WISE is advantageous for a kinematic analysis of complex jet structures having a smooth brightness distribution. As already discussed in Section~\ref{sectwise}, thanks to the long length-chain of seven successive epochs and the symmetric search window used for the VLBA data, our results shown in Figure~\ref{wisel} are not biased and would have a fairly small probability of false detection.



In Figure~\ref{wiselvel}, we present the apparent speeds of the velocity vectors at the finest scale as a function of distance. There is significant dispersion of the speeds at given distances, as already noted in our KaVA analysis and in previous studies (see Section~\ref{sect44}), for the whole distance range. In general, the speeds appear to increase with increasing distance. We present the general trend by using un-weighted binned data points with a bin size of 20 mas, shown by the solid line. They show that the overall jet apparent speeds are larger than $\approx1c$ at $\gtrsim220$ mas and even increase to $\approx5c$ at $\approx380$ mas. This trend is also consistent with our {\tt modelfit} results, demonstrating the reliability of the results obtained by both analysis.

The apparent jet speeds obtained in our study are consistent with those reported by \cite{Asada2014} at distances of $\approx340-410$ mas (see Figure~\ref{wiselvel}). Thus, we confirm the presence of superluminal motions in this region, by using data that is more densely sampled, observed in many more epochs and with a higher angular resolution as compared to the previous study \citep{Asada2014}. However, their data points at $\lesssim320$ mas are generally slower than the overall trend that we found from the WISE analysis, although they are consistent within $1-3\sigma$. Our results indicate that the velocity field at large distances is complicated, such that both fast and slow motions exist. It appears that \cite{Asada2014} could not detect fast motions in some distance ranges at $\lesssim320$ mas probably because of the limited angular resolution and the large monitoring interval of their observations, or the over-simplification of describing the jet emission between $220-320$ mas with three circular Gaussian components.

\section{Discussion}
\label{sect6}

\subsection{Jet re-brightening and apparent stationary features}
\label{discuss1}

It is well known that the M87 jet brightness generally decreases with increasing distance from the core in the region of our interest (at distances less than $\approx400$ mas). However, we found that the jet is re-brightened at $\approx20$ mas in our KaVA data as well as at $\approx65$ and $\approx165$ mas in our VLBA data. The re-brightening in those regions was already shown in previous VLBI observations (e.g., \citealt{Reid1989, Dodson2006, Asada2014}), and those regions were originally labeled as components N2, N1, and M, respectively \citep{Reid1989}. Sub-luminal or quasi-stationary motions were reported in those regions. However, our KaVA observations suggested that the 20 mas feature may not actually be stationary but shows a complicated evolution (Figure~\ref{cleank}). We found that the brightness centroid in the south jet limb keeps moving back and forth. \cite{Dodson2006} pointed out that back and forth motions in the re-brightened regions could be explained if jet components move through standing recollimation shocks, which can naturally form in supersonic jets having a pressure mismatch with an external confining medium (e.g., \citealt{Sanders1983, WF1985, DM1988, Gomez1995, KF1997, Cawthorne2013, Mizuno2015, Fromm2016, Fuentes2018, Park2018}). Indeed, both our {\tt modelfit} analysis (Section~\ref{point}) and WISE analysis (Section~\ref{sectwise}) suggest that the underlying flow speed would actually be quite fast ($\beta_{\rm app}\approx1-3c$) in the 20 mas region.

Our WISE analysis of the VLBA data also show the presence of fast motions at apparent speeds up to $\approx2c$ near the re-brightened region at $\approx165$ mas (see Figures~\ref{wisel} and~\ref{wiselvel}). We note that the results at $\lesssim200$ mas can be affected by the limited angular resolution of our data as we could not resolve the north and south jet limbs. Nevertheless, no slow motion is detected in this region, while fast outward motions of $\beta_{\rm app}\approx2c$ are consistently observed. This implies that the underlying flow speed can be quite fast in this apparently stationary feature, similar to the case of the $\approx20$ mas region. If this is true, then the apparent stationary features originate from the re-brightening of the jet, not from physically standing features within the jet. Future VLBI monitoring observations which can resolve the re-brightened regions will be helpful for confirming this scenario. Also, spectral index and linear polarization analysis will make it clear whether the re-brightening is due to efficient acceleration of synchrotron emitting particles by e.g., recollimation shocks, or due to locally enhanced Doppler boosting through helical motions within the jet.

\subsection{Slow Jet Acceleration\label{sect61}}

\begin{figure*}[t!]
\begin{center}
\includegraphics[trim=4mm 4mm 14mm 2mm, clip, width = 0.78\textwidth]{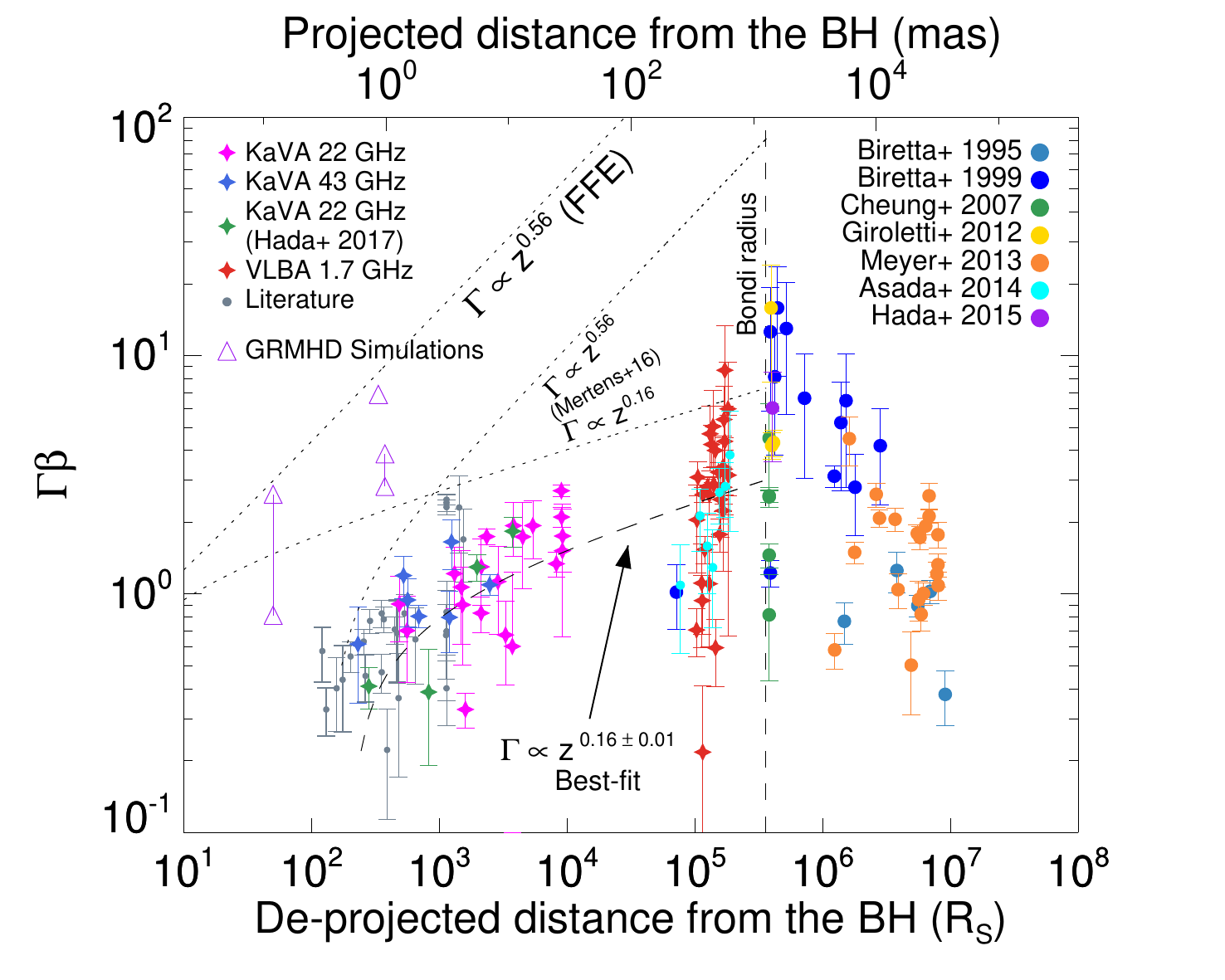}
\caption{Magnitude of the spatial component of the four-velocity $\Gamma\beta$, obtained from the measured apparent speeds with the adopted jet viewing angle of $17^\circ$, as a function of de-projected distance from the black hole in units of $R_{\rm S}$. The data points obtained in this study and our previous study using the KaVA observations at 22 GHz \citep{Hada2017} are shown with the diamonds, while those obtained in the literature presented in the right panel of Figure~\ref{vel} are shown with the grey filled circles. We include the data points in the literature for the jet speeds on large scales as well \citep{Biretta1995, Biretta1999, Cheung2007, Giroletti2012, Meyer2013, Asada2014, Hada2015}. The distance between the radio core and the black hole are corrected by using the core-shift measurement \citep{Hada2011}. We also include the results obtained by GRMHD simulations shown with the purple open upward triangles \citep{McKinney2006, Penna2013, Nakamura2018}. The two triangles connected with the vertical solid lines show the range of speeds obtained by using different assumed black hole spins in the simulations. The best-fit function, assuming a power-law function for the bulk Lorentz factor and converting it into the four-velocity, to the data points obtained in this study and our previous study (the magenta, blue, red, green diamonds) is $\Gamma \propto z^{0.16\pm0.01}$ and shown with the black dashed line. Some data points showing slow or quasi-stationary motions are not included in the fitting (see Section~\ref{sect61} for more details). The linear jet acceleration profile of $\Gamma\propto z^{0.56}$, expected in the FFE model (see texts) for the assumed black hole spin of $a=0.9$, and the upper envelopes of $\Gamma\propto z^{0.56}$ and $\Gamma\propto z^{0.16}$ presented in \cite{Mertens2016}, are shown as a reference with the black dotted line. Note that the power-law profiles for $\Gamma$ are shown by the curved lines in the logarithmic plot of $\Gamma\beta$ versus distance.  \label{lorentz}}
\end{center}
\end{figure*}

In Figure~\ref{lorentz}, we present the magnitude of the spatial component of the four-velocity $\Gamma\beta$, where $\Gamma$ is the bulk Lorentz factor, as a function of de-projected distance from the black hole in units of $R_{\rm S}$. We note that we do not include the VLBA data points at $\lesssim200$ mas detected by the WISE analysis, which can be affected by the limited angular resolution and by the re-brightened jet regions, to be conservative. We corrected for the core positions with respect to the jet base by using the core-shift measurements by \citealt{Hada2011}. We include four data points from our recent study of KaVA monitoring data observed in 2013--2014 at 22 GHz \citep{Hada2017}. We also include the results obtained in the literature to compare with our results and to show an overall trend of jet acceleration and deceleration at distances from sub-pc to kpc scales. We discuss several implications of our results below.

One of the notable features we found is that the jet is moving at subluminal speeds at distances $\lesssim1$ mas, corresponding to de-projected distances $\lesssim500\ R_{\rm S}$. This is consistent with the results of other studies, especially with the speeds obtained by the jet to counterjet brightness ratio (see Section~\ref{sect44} and the right panel of Figure~\ref{vel}). This result indicates that it is less likely that VLBI observations are missing very fast jet motions due to the limited angular resolution or cadence on this scale. The jet becomes relativistic at a distance $\approx10^3R_{\rm S}$, although it is difficult to determine an exact location because of the dispersion in the observed speeds. This distance is relatively far from the central engine compared to the results of various GRMHD simulations which obtain $\Gamma\gtrsim$ a few already at distances less than a few hundred $R_{\rm S}$ (e.g., \citealt{McKinney2006, Penna2013, Nakamura2018}, see also Figure~\ref{lorentz}).

In addition, the observed jet acceleration profile seems to be relatively flat compared to the prediction of magnetic jet acceleration models. To describe a general trend of jet acceleration, we fit a function, assuming a power-law function for the bulk Lorentz factor and converting it into the four-velocity, to the data points obtained in our study and our previous study \citep{Hada2017}. We did not include the KaVA data points showing negative or zero speeds (7 out of 39) and the VLBA data points showing $\beta_{\rm app} \leq1c$ (11 out of 27) for the fitting because those data points cannot properly represent the general acceleration trend. Also, there is a possibility that those slow motions are associated with instability patterns, not with the bulk jet speed (e.g., \citealt{Lobanov2003, HE2011, Mertens2016}, see Section~\ref{sect62} for related discussions). We obtain the best fit of $\Gamma \propto z^{0.16\pm0.01}$, which is shown by the dashed line in Figure~\ref{lorentz}.

We note, however, that this profile assumes that all the data points follow the same power-law, which may not be necessarily the case (see Section~\ref{sect62}). Theoretical studies of highly magnetized jets in the highly relativistic limit ($\sigma \gg 1$, where $\sigma$ is the Poynting flux per unit matter energy flux, so-called the magnetization parameter) or in the far zone ($r \gg r_{\rm lc}$, where $r_{\rm lc} = c/\Omega$ is the light cylinder radius with $\Omega$ being the angular velocity of a given streamline) show that an evolution of the Lorentz factor would be described as $\Gamma \propto R \propto z^a$ near the jet base (so-called the linear acceleration regime) with a transition to a slower acceleration profile at a certain distance (e.g., \citealt{Tchekhovskoy2008, Komissarov2009, Lyubarsky2009}). The latter proportionality comes from the jet collimation profile, which was found to be $a \approx0.56$ for M87 in the regions we are probing in this study \citep{AN2012, Hada2013, NA2013}. Thus, the observed trend of jet acceleration appears to be much flatter than the linear acceleration profile of $\Gamma \propto z^{0.56}$.

Taken as a whole, our results suggest that the M87 jet is gradually accelerated over a large jet distance range that coincides with the jet collimation zone, which is one of the essential characteristics of the magnetic jet acceleration mechanism (e.g., \citealt{VK2004, Lyubarsky2009}). However, the observed acceleration does not seem to be as efficient as in the models or in the results of GRMHD simulations of a highly magnetized jet. There are two essential ingredients necessary for efficient jet acceleration in the models: the degree of jet magnetization near the jet base (often given by $\sigma$) and the ``differential collimation'' of poloidal magnetic fields. The former tells us about the amount of electromagnetic energy available for being converted into the jet kinetic energy and thus determines the upper limit of jet bulk Lorentz factor. The latter is realized when the inner streamlines closer to the jet axis are more collimated than the outer ones, also known as the ``magnetic nozzle'' effect (e.g., \citealt{Li1992, VK2003}), and it determines the efficiency of conversion from electromagnetic to kinetic energy.

Therefore, the observed slow jet acceleration may be explained if the M87 jet is not highly magnetized at its base. \cite{Mertens2016} have found a transition from an efficient linear acceleration ($\Gamma \propto z^{0.56}$) to a slower acceleration ($\Gamma \propto z^{0.16}$) occurring at $\approx10^3\ R_{\rm S}$. The latter profile at outer jet distances is in good agreement with the acceleration profile we found, while the former at inner distances is much steeper. This discrepancy might be related with the fact that they used the fastest jet motions at a given distance bin for deriving the profiles (see Section~\ref{sect62} for related discussions). They applied the asymptotic solution of relativistic, axisymmetric MHD equations in the far zone derived by \cite{Lyubarsky2009} for the case of the pressure of an external confining medium ($P_{\rm ext}$) rapidly decreasing with distance from the central engine, i.e., $\kappa>2$ in $P_{\rm ext} \propto z^{-\kappa}$. This solution predicts a transition of jet acceleration and collimation profiles from the linear acceleration ($\Gamma \propto R \propto z^{\kappa/4}$) with a parabolic jet shape to a slower acceleration ($\Gamma \propto z^{(\kappa-2)/2}$) with a conical jet shape. While \cite{Mertens2016} obtained a good fit with the inferred value of $\kappa\approx2.4$ using this model, a transition to the conical jet shape was not found in the region they probed\footnote{We note that our recent study of Faraday rotation in the jet at distances $\lesssim2\times10^5\ R_{\rm S}$ suggests $\kappa\lesssim2$ \citep{Park2019}, which allows a parabolic jet shape without a transition to a conical shape \citep{Komissarov2009}, as observed in this region.} \citep{AN2012}. They explained this contradiction with early saturation of the Poynting flux, resulting in a quenched acceleration. If this is the case, the assumptions of a Poynting flux dominated jet, i.e., $\sigma\gg1$, used in the models may not hold. The jet may have a relatively small initial magnetization parameter if it is launched in the inner part of the accretion disk\footnote{We note, however, that GRMHD simulations consistently found that gas outflows launched from the disk cannot reach relativistic speeds due to high mass-loading (e.g., \citealt{Sadowski2013, Yuan2015, Nakamura2018, Qian2018}).} (e.g., \citealt{Mertens2016, Kim2018}).

In contrast, previous studies constrained the degree of magnetization near the jet base of M87, based on VLBI observations \citep{Kino2014, Kino2015b, Kim2018}. They suggested that the jet base is highly magnetized, which is in line with indirect observational evidence that M87 is in a magnetically arrested disk (MAD, \citealt{Narayan2003, Tchekhovskoy2011}) state (e.g., \citealt{Zamaninasab2014, Park2019}). If this is the case, the distance at which the jet transitions from the linear acceleration regime to a slower acceleration regime would be given by steady axisymmetric force-free electrodynamic (FFE) solutions \citep{Tchekhovskoy2008}:
\begin{equation}
z_{\rm tr} = z_{\rm fp}\left[\frac{1}{\Omega_{\rm fp} z_{\rm fp}}\frac{C}{2\sin^2(\theta_{\rm fp}/2)\sqrt{(2-\nu)\nu}}\right]^{1/(1-\nu)},
\end{equation}

\noindent where $z_{\rm fp}$, $\Omega_{\rm fp}$, and $\theta_{\rm fp}$ are the distance from the central engine, the rotational frequency, and the colatitude angle at the footpoint of the local field line, respectively. $C$ is a numerical factor that depends on the field line rotational profile, $\nu$ is the radial power-law index in the poloidal flux function of the initial magnetic field configuration which describes the asymptotic shape of the field line as $z\propto R^{2/(2-\nu)}$. Recent GRMHD simulations found that the jet collimation profile of M87 is in good agreement with the outermost parabolic streamline of the FFE solution anchored to the black hole event horizon on the equatorial plane \citep{Nakamura2018}. Based on this result, one may use $\theta_{\rm fp} = \pi/2$, $\nu = 0.89$, and $C = \sqrt{3}$. $\Omega_{\rm fp}$ and $z_{\rm fp}$ depend on the black hole spin which was estimated to be $|a|\gtrsim0.2$ for M87 (\citealt{Doeleman2012, EHT2019e}, see also \citealt{Nokhrina2019}). Thus, the expected transition distance is $z_{\rm tr} \gtrsim 2.5\times10^7\ R_{\rm S}$, indicating that the linear acceleration profile of $\Gamma \propto z^{0.56}$ would be maintained in the observed jet acceleration zone according to the FFE model.

Therefore, the observed trend of jet acceleration seems difficult to explain with the FFE model. It may indicate that, if the jet is initially highly magnetized as previous studies have suggested, there may be a lack of differential collimation of poloidal fields in the M87 jet. Indeed, recent GRMHD simulations showed that the differential collimation proceeds in a complicated manner, depending on the distance from the central engine and the black hole spin \citep{Nakamura2018}. This indicates that an efficient jet acceleration through the Poynting flux conversion predicted in the FFE models may not be always achieved. In this case, the jet would still remain Poynting flux dominated even beyond the acceleration and collimation zone\footnote{However, there is indication that the energy density of relativistic electrons is larger than that of magnetic fields by orders of magnitude in the jets of several blazars (e.g., \citealt{Kino2002, Hayashida2015}), for which most radio emission is thought to come from beyond the acceleration and collimation zone (e.g., \citealt{Marscher2008}).} (outside the location of HST-1) because not all of the Poynting flux would be converted into the kinetic energy\footnote{unless much of the jet electromagnetic energy is dissipated into other forms of energy instead of being transferred into the jet kinetic energy (e.g., \citealt{Ostrowski1998, SO2002, Giannios2009, SS2014})}. We note that several studies have pointed out that the M87 jet may be highly magnetized on kpc scales from the observed morphology and linear polarization structure (e.g., \citealt{Owen1989}), the high energy $\gamma$-ray observations \citep{Stawarz2005}, and the conical jet expansion observed in the region where a surrounding interstellar medium is nearly uniformly distributed \citep{AN2012}.

In our analysis and discussion above, we assumed that the jet viewing angle is constant over the distance range of our interest. However, one can see that local changes of the direction of the jet ridge on the sky plane are present in the VLBA images (Figure~\ref{vlbaclean}). Also, the jet opening angle decreases with increasing distance (e.g., \citealt{Junor1999, AN2012}), which can change the effective jet viewing angles at different distances. These effects may contribute to the observed jet acceleration profile and the scatter of data points (Section~\ref{sect62}). However, the viewing angle constraints on mas scales (e.g., \citealt{Mertens2016}) and arcsecond scales \citep{Biretta1999} are consistent with each other. Furthermore, the VLBA images in Figure~\ref{vlbaclean} suggest that the jet morphology is globally straight on the sky plane at $\lesssim 420$ mas, indicating that the variation of the jet viewing angle with jet distance would not be significant. Therefore, we expect that the observed jet acceleration profile would not be much affected by the assumption of constant jet viewing angle, although quantitative examination is needed in future studies.

\subsection{Multiple Streamlines and Velocity Stratification\label{sect62}}

The above discussion is based on the best-fit function of a simple power-law for the bulk Lorentz factor, which may hold only for a single streamline in the FFE models. However, it is possible that the observed jet emission consists of multiple streamlines. In this case, a more complicated jet velocity field with a signature of velocity stratification is expected. Different streamlines may have different magnetization parameters (e.g., \citealt{TT2003}) and different collimation profiles (e.g., \citealt{Komissarov2007}), which can result in a lateral stratification in jet velocity. The distribution of electric current flows within the jet, which is likely associated with the rotational velocity profile of the footpoint of jet (e.g., \citealt{Komissarov2007, Tchekhovskoy2008, Komissarov2009}), determines which streamlines are more efficiently accelerated than other streamlines. In the case of black-hole driven jets, the outer jet layers become faster than the inner layers, which may be associated with the limb-brightend feature seen in M87. This is recently suggested by a semi-analytic model \citep{Takahashi2018} and GRMHD simulations \cite{Nakamura2018}. We note that additional mass-loading of the jet by the interaction with the confining wind could also contribute to the velocity stratification (e.g., \citealt{Chatterjee2019}).

Indeed, we found the presence of dispersion in the observed speeds at a given distance in both our KaVA and VLBA data (see Figure~\ref{lorentz}). If the dispersion originates from different streamlines having different speeds at the same distance, then it is difficult to discuss the efficiency of Poynting flux conversion by comparing with the bulk Lorentz factor profiles prediced by the FFE models, which are given by a single power-law for a single streamline. We note that the different speeds observed at HST-1 between optical \citep{Biretta1999} and radio wavelengths (\citealt{Cheung2007, Giroletti2012, Hada2015}, see also Figure~\ref{lorentz}) may originate from jet emission at different frequencies dominated by different jet layers (e.g., \citealt{Mertens2016, Kim2018, Walker2018}). This scenario may also be supported by observations of different jet widths and linear polarization structure between radio and optical on kpc scales (e.g., \citealt{Sparks1996, Perlman1999})

Alternatively, the fastest motions at a given distance may represent the jet bulk motions, while slower motions are associated with instability patterns or outer winds moving at sub-relativistic speeds launched by the accretion disk, as suggested by \cite{Mertens2016}. In this scenario, the observed jet emission consists of a single streamline. They selected the fastest 10\% of the speeds measured within individual distance bins, and found a transition from an efficient linear acceleration ($\Gamma \propto z^{0.56}$) to a slower acceleration ($\Gamma \propto z^{0.16}$) occurring at $\approx10^3\ R_{\rm S}$. The upper envelopes of speeds used in their study, shown with the dotted lines in Figure~\ref{lorentz}, seem to be consistent with our data points as well. \cite{Mertens2016} suggested that the presence of a transition may imply that the jet is not initially highly magnetized and early saturation of the Poynting flux conversion results in a quenched acceleration, as discussed in Section~\ref{sect61}. The slow inward motions detected by our WISE analysis (see Figures~\ref{vel} and ~\ref{wiselvel}) may be in line with this scenario. Those motions could be produced by instabilities and/or turbulence in the jet.

It is difficult to determine whether the observed velocity dispersion originates from (i) multiple streamlines following different acceleration profiles or (ii) fast bulk jet motions and slow outer winds/instability patterns with the data presented in this study. These scenarios can be tested by observing a systematic acceleration of the \emph{lower envelope} of the velocity field, which would be difficult to be reproduced by instability patterns (e.g., \citealt{Hardee2000, Lobanov2003, HE2011}) or winds launched from hot accretion flows (e.g., \citealt{Yuan2015}). We note that the fraction of relativistic speeds seems to be higher at larger distances, e.g., 7/15 and 9/12 for distances of 200--300 and 300--400, respectively (Figure~\ref{wiselvel}), although the small number of data points prevents us from confirming the existence of the lower envelope acceleration.

\subsection{Current Limitations and Future Prospects\label{sect63}}

We remark on the limitations of our present study and address the need for future studies. As was explained in Section~\ref{sect44}, we applied various methods for the jet kinematics because each method has its own advantages and disadvantages. Therefore, the dispersion in the observed speeds at a given distance bin (the left panel of Figure~\ref{vel}) may simply arise due to potential systematic errors in the different methods. However, the dispersion may not be solely due to these errors because many previous studies which use a single kinematic method have shown non-negligible dispersion at a given distance bin (see, e.g., the right panel of Figure~\ref{vel} and references therein). Although we see an indication of underlying fast flows in the re-brightened jet regions, this needs to be confirmed with higher resolution data. Also, it is necessary to perform a jet kinematic analysis for the distance range of $10^4-10^5\ R_{\rm S}$ with higher resolution data to fill in the gap in the current velocity field diagram. The above limitations require dedicated monitoring observations with a high angular resolution, sensitivity, and an observing cadence in the future. On-going and future monitoring observations with the EAVN will be important in this regard (Y. Cui et al. 2019, in preparation).

\section{Conclusions}
\label{sect7}

We have studied the kinematics of the M87 jet with KaVA monitoring observations performed in eight epochs in 2016 quasi-simultaneously at 22 and 43 GHz. We also performed a complementary kinematic analysis of the VLBA archive data observed in 19 epochs between 2005 and 2009 at 1.7 GHz. Our work leads us to the following principal conclusions:

\begin{enumerate}

\item We found that the apparent jet speeds generally increase from $\approx0.3c$ at $\approx0.5$ mas from the core to $\approx2.7c$ at $\approx20$ mas, which indicates that the jet is accelerated from subluminal to superluminal speeds on this scale, as recent studies have suggested \citep{Mertens2016, Hada2017, Walker2018}.

\item We found that the jet moves at relativistic apparent speeds up to $\approx5.8c$ at distances $\approx200-410$ mas. Combined with the kinematic results for the inner jet regions, the M87 jet seems to gradually accelerate over a broad distance range from $10^2$ to a few $\times10^5\ R_{\rm S}$, while it is being gradually collimated simultaneously, as the magnetic jet acceleration models predict.

\item We observe relativistic jet motions at speeds of up to $\approx2.7c$ in the apparent stationary features at distances of $\approx20$ and 165 mas. This indicates that those features may not be physically stationary but result from the re-brightening of the jet.

\item Both jet kinematic analysis using VLBI monitoring observations and an analysis of the brightness ratio of the jet and counterjet suggest that the jet is moving at subluminal speeds at de-projected distances $\lesssim500\ R_{\rm S}$. This result indicates that the jet is in a non-relativistic regime up to distances considerably larger than what previous GRMHD simulations predicted.

\item We first compared the observed trend of jet acceleration with the models of highly magnetized jets, based on the assumption that the jet emission consists of a single streamline. We described the observed jet speeds at distances of $\approx600-200,000\ R_{\rm S}$ with a simple power-law function for the bulk Lorentz factor and obtained $\Gamma \propto z^{0.16\pm0.01}$. This profile is much flatter than the model prediction. This result indicates that the jet is not highly magnetized near its base and early saturation of Poynting flux leads to the flat acceleration profile, or the jet is highly magnetized but Poynting flux conversion through the differential collimation of poloidal magnetic fields in the jet may not be very efficient.

\item However, we found that there is a non-negligible dispersion in the observed speeds at a given distance both at $\lesssim20$ mas and $\approx200-410$ mas. This suggests that the jet velocity field of M87 may be stratified. In this case, the above interpretation based on a simple power-law fitting could be oversimplified. The dispersion may originate from either multiple streamlines following different acceleration profiles in the jet or from faster bulk jet motions and slower outer winds/instability patterns. Investigating whether there is a systematic acceleration of the lower envelope of the jet speeds with future VLBI observations will be important to test these scenarios.

\end{enumerate}

We finally remark on that the results presented in this paper are derived from the radial kinematic analysis using the first year data observed with KaVA in our large program. Different types of analysis such as the spectral evolution between 22 and 43 GHz and jet motions in the transverse direction to the jet axis using these data will be presented elsewhere (H. Ro, et al. 2019, in preparation). Furthermore, our program has begun using the EAVN extensively since 2017 with shorter intervals down to $\sim5$ days for specific periods, which allows us to obtain high-quality images of the jet extended down to $\sim30$ and $\sim10$ mas at 22 and 43 GHz, respectively. The results using these data will be presented in forthcoming papers (Y. Cui, et al. 2019, in preparation). Thanks to the advent of millimeter VLBI arrays such as the EHT (e.g., \citealt{EHT2019a}) and the global millimeter VLBI array (GMVA; e.g., \citealt{Hodgson2017, Kim2018}), resolving the horizon-scale structure of accreting and outflowing matters has been realized. We stress that continued monitoring observations with centimeter VLBI arrays such as the EAVN in synergy with mm-VLBI observations will contribute to complete the picture of jet launching, acceleration, and collimation.

\acknowledgments 
We thank the referee for constructive comments, which helped improve the paper. We acknowledge all staff members and students at KVN and VERA who supported the operation of the array and the correlation of the data. J.P. thanks K. Toma for helpful discussions. KVN is a facility operated at by the Korea Astronomy and Space Science Institute. VERA is a facility operated at National Astronomical Observatory of Japan in collaboration with associated universities in Japan. The Very Long Baseline Array is an instrument of the Long Baseline Observatory. The Long Baseline Observatory is a facility of the National Science Foundation operated by Associated Universities, Inc. We acknowledge financial support from the Korean National Research Foundation (NRF) via Global Ph.D. Fellowship Grant 2014H1A2A1018695 (J.P.), 2015H1D3A1066561 (I.C.), and Basic Research Grant NRF-2015R1D1A1A01056807 (S.T.). This work by Jeffrey A. Hodgson was supported by Korea Research Fellowship Program through the National Research Foundation of Korea (NRF) funded by the Ministry of Science and ICT(2018H1D3A1A02032824). For this research K.H. and F.T. are supported by the Japan Society of the Promotion of  Science (JSPS) KAKENHI grant number 18KK0090. M.K. acknowledges the financial support of JSPS KAKENHI program with the grant numbers of JP18K03656 and JP18H03721. T.J. and G.-Y.Z. are supported by Korea Research Fellowship Program through the National Research Foundation of Korea (NRF) funded by the Ministry of Science, ICT and Future Planning (NRF-2015H1D3A1066561). Part of this work was achieved using the grant of Visiting Scholar Program supported by the Research Coordination Committee, National Astronomical Observatory of Japan (NAOJ). J.-C.A acknowledges support from the Malaysian Fundamental Research Grant Scheme (FRGS) FRGS/1/2019/STG02/UM/02/6


\begin{appendix}

\section{WISE Analysis of KaVA data}
\label{kavawise}

\begin{figure*}[t!]
\begin{center}
\includegraphics[trim=0mm 0mm 0mm 0mm, clip, width = 0.8\textwidth]{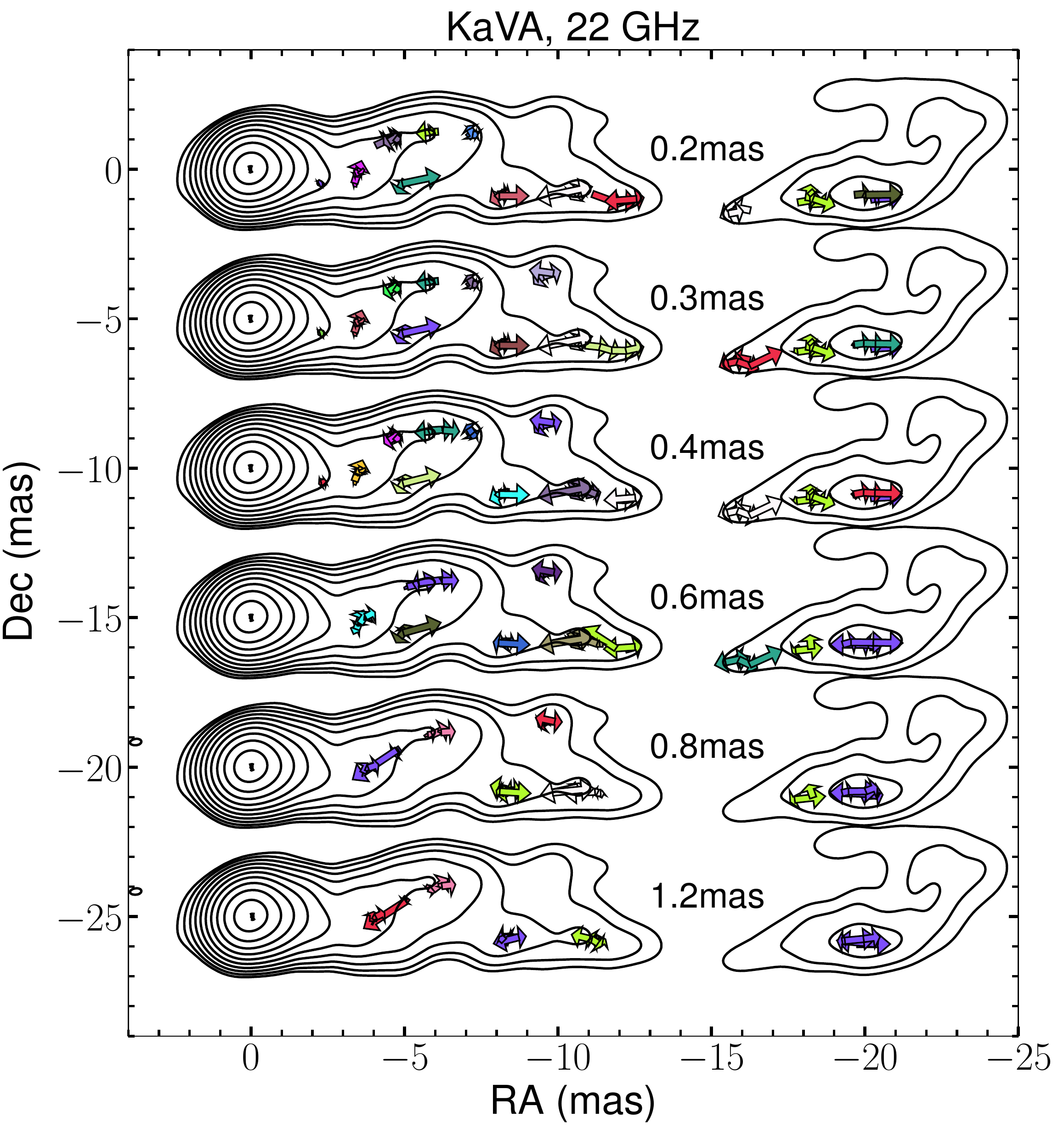}
\caption{Same as Figure~\ref{wise} but showing whole displacement vectors of the SSPs detected by WISE for the KaVA data at 22 GHz. The results for six different SWD/IWD scales are shown in the stacked CLEAN maps shifted along the y-axis.
\label{wisekcol}}
\end{center}
\end{figure*}

\begin{figure*}[t!]
\begin{center}
\includegraphics[trim=0mm 0mm 0mm 0mm, clip, width = 0.44\textwidth]{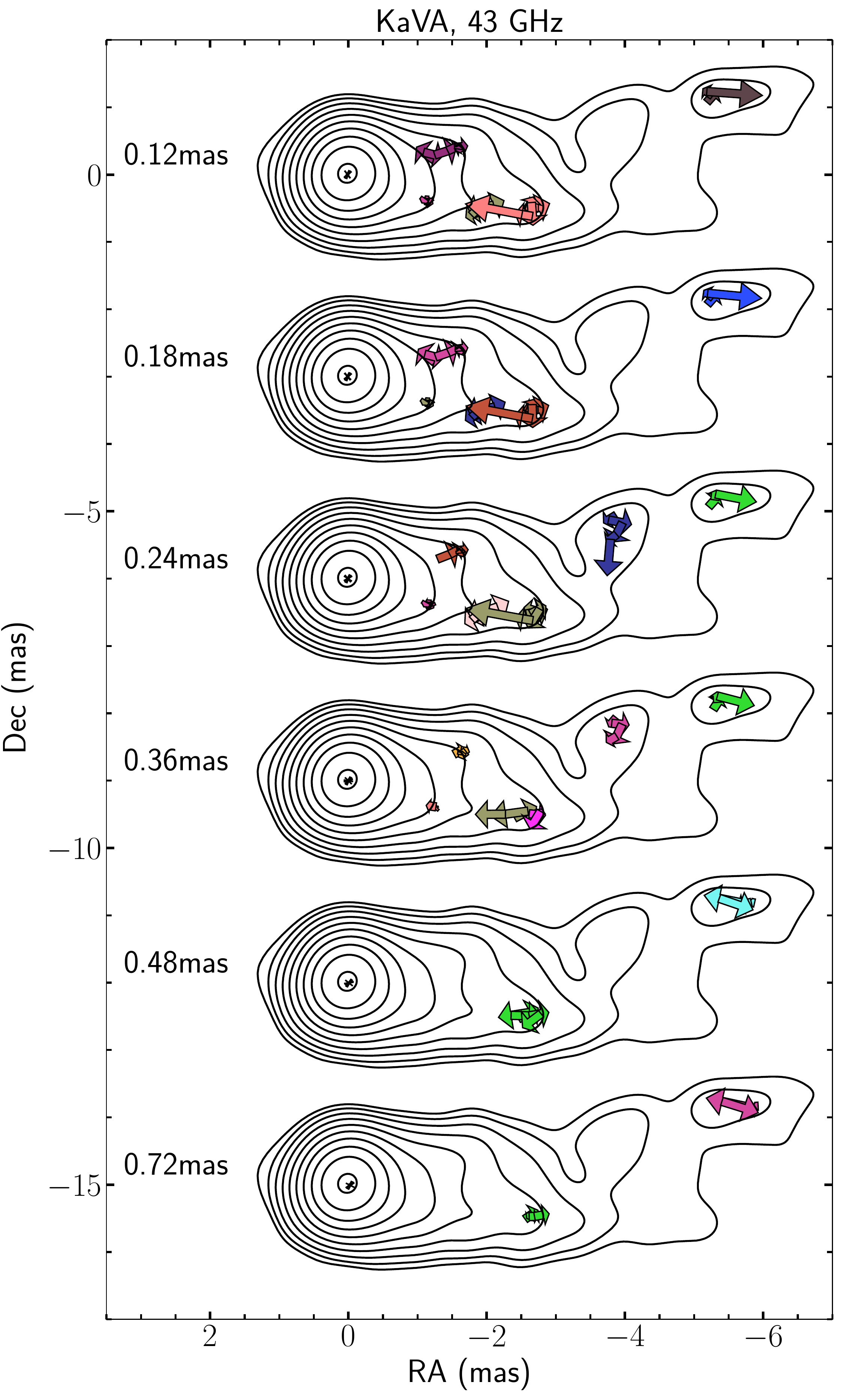}
\caption{Same as Figure~\ref{wisekcol} but for the KaVA data at 43 GHz.
\label{wiseqcol}}
\end{center}
\end{figure*}

\begin{figure*}[t!]
\begin{center}
\includegraphics[trim=0mm 0mm 0mm 0mm, clip, width = 0.48\textwidth]{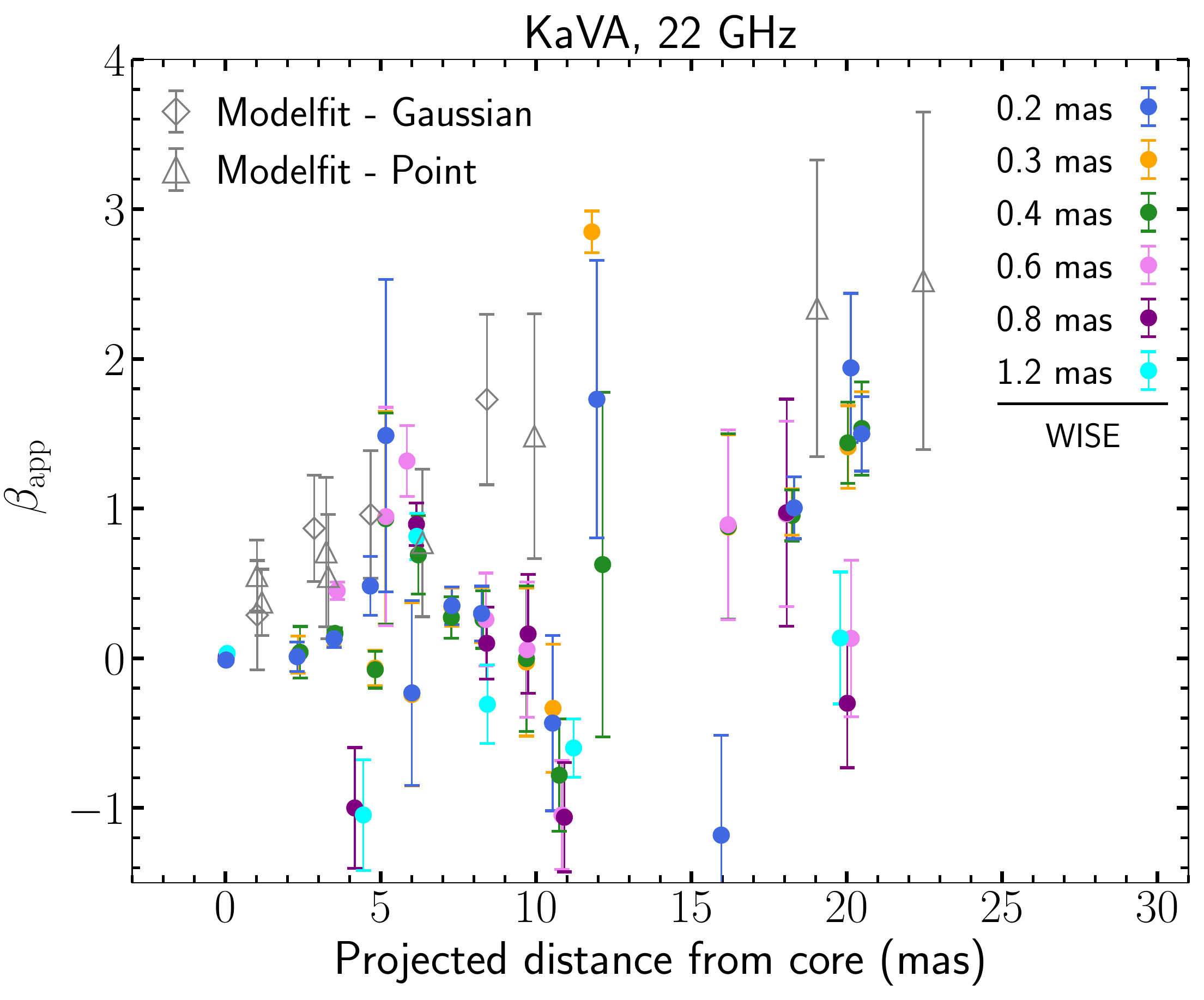}
\includegraphics[trim=0mm 0mm 0mm 0mm, clip, width = 0.51\textwidth]{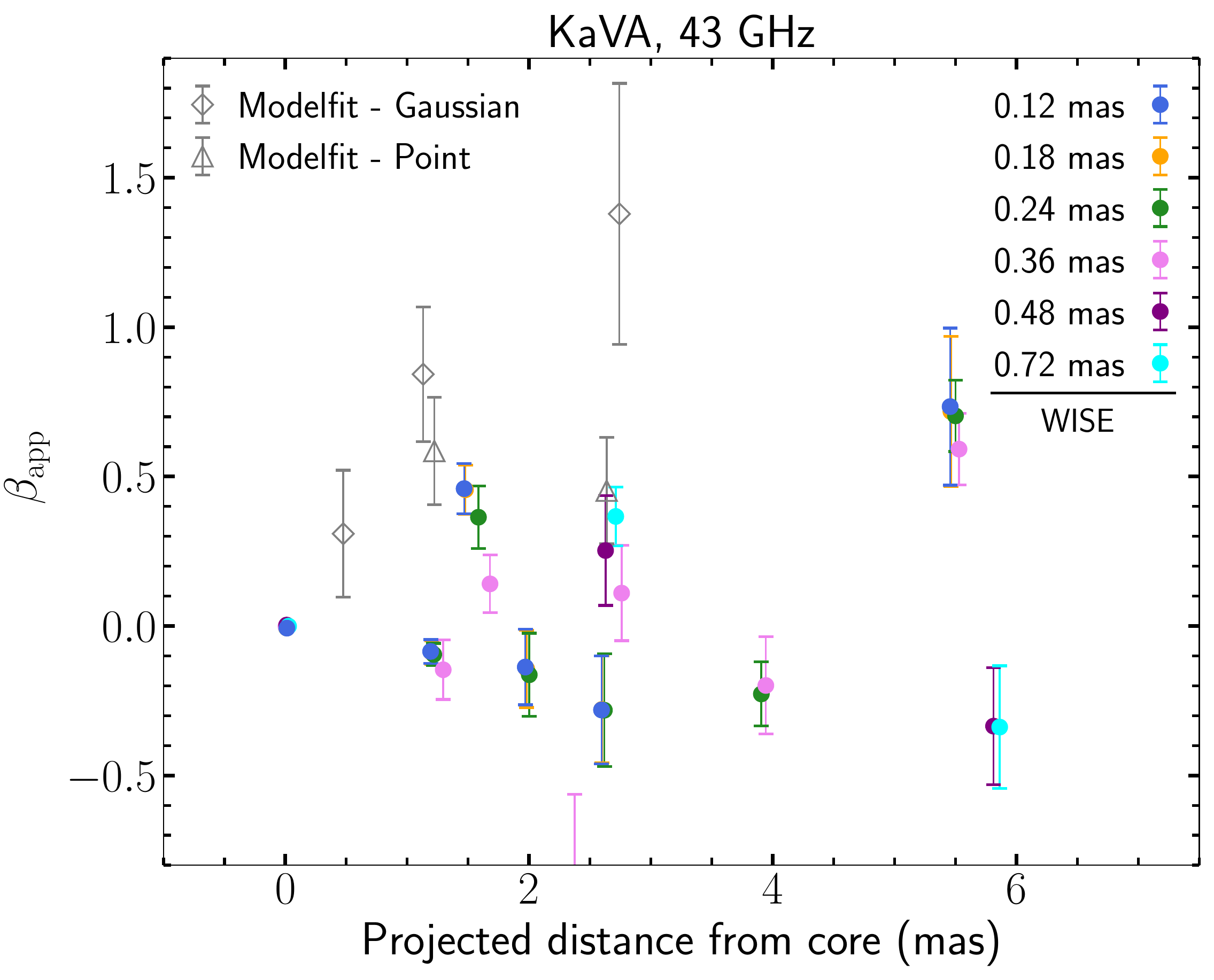}
\caption{Apparent speeds in units of the speed of light as functions of projected distance from the core in units of mas detected by the WISE analysis of the KaVA data (filled circles) on six different SWD/IWD scales (shown in different colors), and detected by the {\tt modelfit} analysis obtained in Section~\ref{circularGaussian} (grey open upward triangles) and~\ref{point} (grey open diamonds) at 22 (left panel) and 43 GHz (right panel). 
\label{wisevel}}
\end{center}
\end{figure*}

In Section~\ref{sectwise}, we present the results of an application of the WISE technique to our KaVA data. In this appendix, we present all of the displacement vectors of SSPs on different SWD/IWD scales.

In Figures~\ref{wisekcol} and~\ref{wiseqcol}, we show the displacement vectors on top of the stacked CLEAN maps at 22 and 43 GHz, respectively. The results for six different scales are displayed on the maps shifted along the y-axis. Some of the displacement vectors obtained on different scales in the same parts of the jet are found to be significantly different from each other. Sometimes, we see gradual outward motions with relatively fast speeds on larger scales, while we see slower or quasi-stationary motions on smaller scales. For example, this behavior is seen at distances $\approx3.5$ and $\approx5$ mas at 22 GHz. On the other hand, sometimes we observe faster motions on smaller scales and slower or quasi-stationary motions on larger scales. For example, this behavior is seen at $\approx1.5$ mas at 43 GHz. The scale dependence becomes more prominent on the largest scales especially at 22 GHz. We see negative motions at fast speeds at $\approx4$ and $\approx10$ mas, while gradual outward motions or stationary motions are observed in the same parts on smaller scales.

Previous studies of the jet kinematics for M87 \citep{Mertens2016} and for the nearby radio galaxy 3C 264 \citep{Boccardi2019} using WISE found that the results on various SWD/IWD scales are consistent with each other. Thus, they selected the result on the finest scale, which provides the highest effective spatial resolution among different scales. However, we found non-negligible scale dependence for our KaVA data. Since our data have a smaller sampling interval ($\approx2$ weeks) compared to the previous studies ($\gtrsim3$ weeks), the limited angular resolution of KaVA could be a reason for the scale dependence. Our future, high-resolution observations with the EAVN will be helpful for confirming this conjecture. Based on this argument, we selected the results of the finest scales as our representative WISE result in Section~\ref{sectwise}.

In Figure~\ref{wisevel}, we present the apparent speeds derived from the detected displacement vectors presented in Figures~\ref{wisekcol} and~\ref{wiseqcol}, together with those obtained in our {\tt modelfit} analysis (Section~\ref{circularGaussian} and~\ref{point}), as functions of projected distance from the core. As already noted in Section~\ref{sect44}, the speeds of the fast motions are consistent with the {\tt modelfit} results, while much slower motions are also detected by WISE.

\section{WISE Analysis of VLBA data}
\label{vlbawise}

\begin{figure*}[t!]
\begin{center}
\includegraphics[trim=0mm 0mm 0mm 0mm, clip, width = 0.495\textwidth]{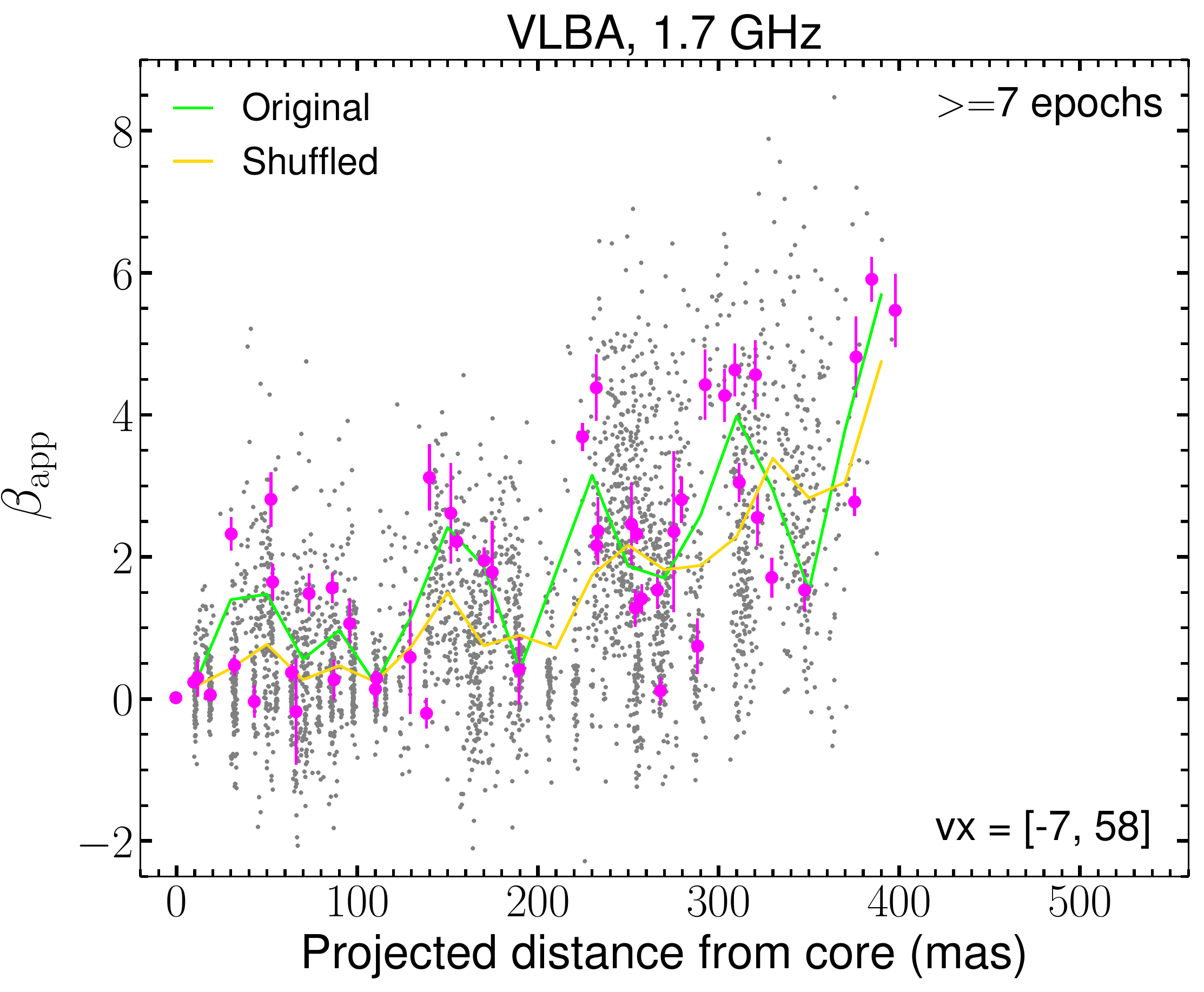}
\includegraphics[trim=0mm 0mm 0mm 0mm, clip, width = 0.495\textwidth]{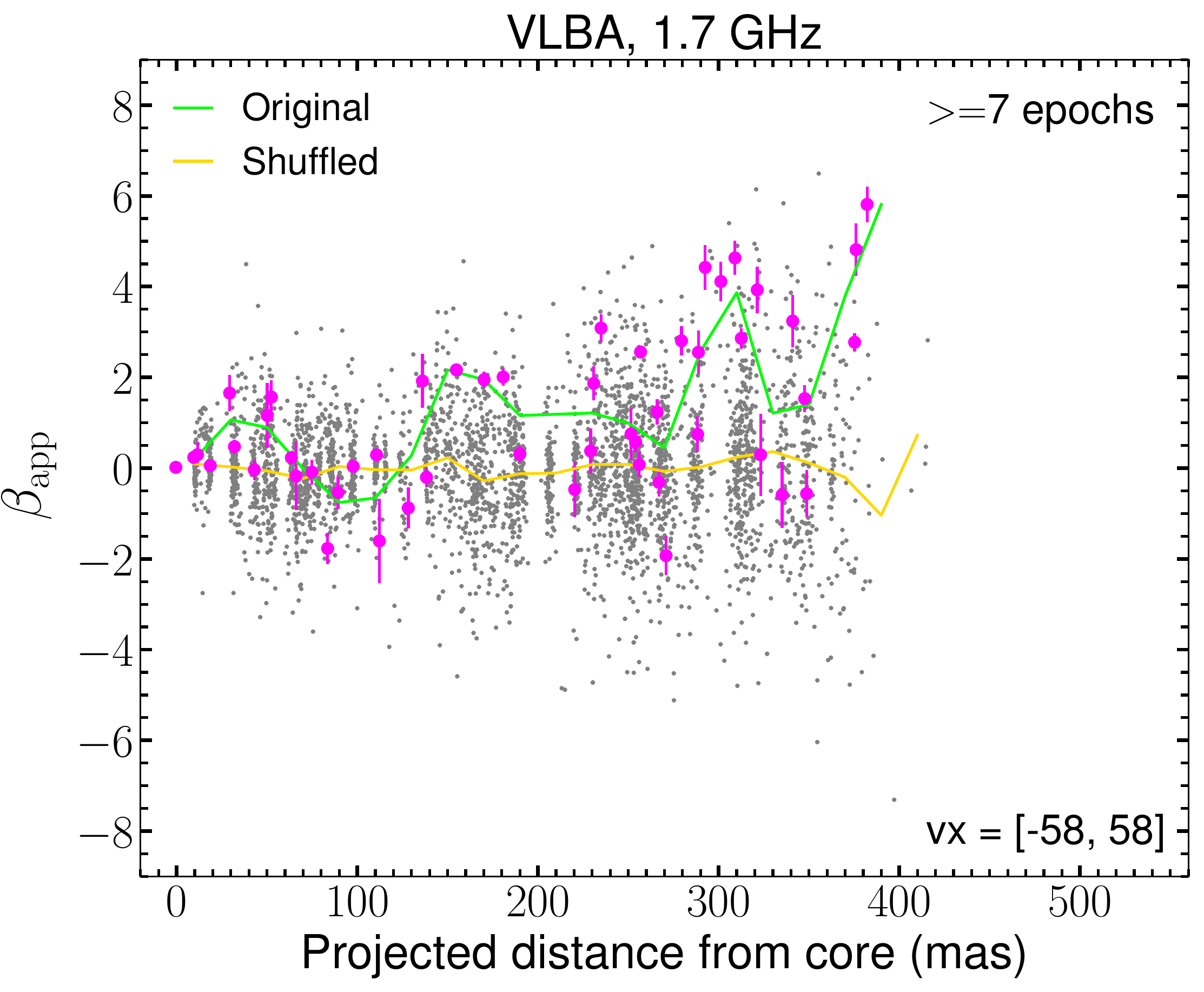}
\caption{Apparent speeds in units of the speed of light as functions of projected distance from the core in units of mas detected by the WISE analysis of the VLBA 1.7 GHz data (magenta filled circles) and of the hundred shuffled data (grey dots) for an asymmetric (left) and symmetric (right) velocity range contraints used for matching SSPs. The speeds for vectors identified in more than seven successive epochs are shown. The green and yellow solid lines represent un-weighted binned data for the original and shuffled data, respectively.
\label{wiselsig}}
\end{center}
\end{figure*}

In this appendix, we present several tests for our WISE results by using the VLBA 1.7 GHz data, which provide us with a number of detected WISE vectors and good statistics. To ensure that our WISE results represent real jet motions instead of random motions produced by potential mis-identification of SSPs in adjacent epochs, we generate one hundred randomly shuffled data sets and perform the WISE analysis with the same parameters used for the original data. In the left panel of Figure~\ref{wiselsig}, we show the apparent speeds of the original data (magenta data points) and the random data set (grey dots) as functions of distance. Interestingly, they show quite similar distributions, which can also be seen by the un-weighted binned data points (the solid lines).

We found that this result is due to the asymmetric velocity constraint along the jet direction applied for the MCC analysis (Section~\ref{sectvlbawise}). The overall distances between matched SSPs in adjacent epochs become larger at larger distances probably because the jet is better resolved and the jet moves at faster speeds. Thus, one can naturally expect a random distribution of apparent speeds with zero means and with its standard deviation becoming larger at larger distances for the shuffled data. However, the asymmetric velocity constraint discards most of the detected negative speeds. Therefore, the overall acceleration trend appears for the shuffled data as well.

\begin{figure*}[t!]
\begin{center}
\includegraphics[trim=0mm 0mm 0mm 0mm, clip, width = 0.495\textwidth]{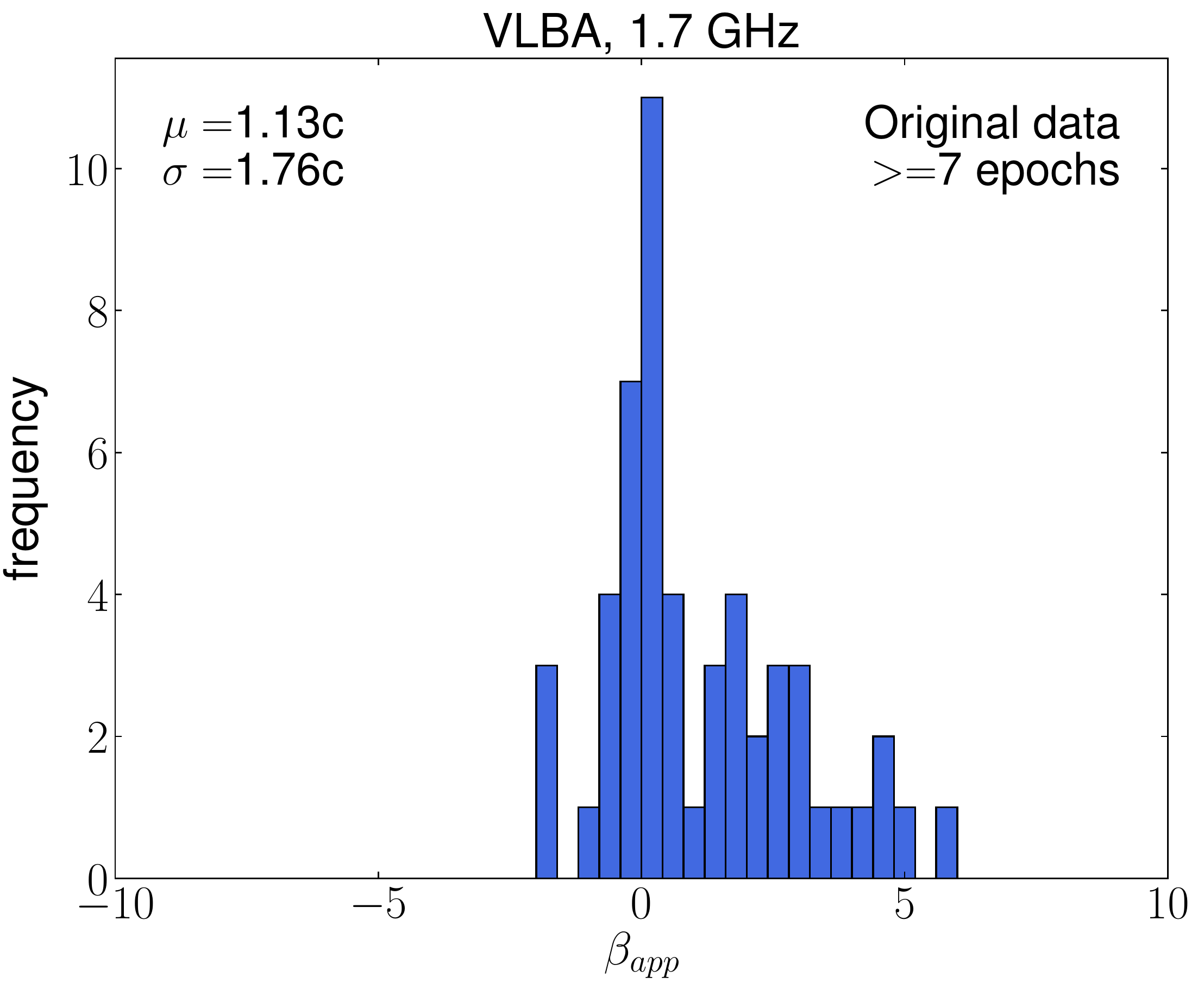}
\includegraphics[trim=0mm 0mm 0mm 0mm, clip, width = 0.495\textwidth]{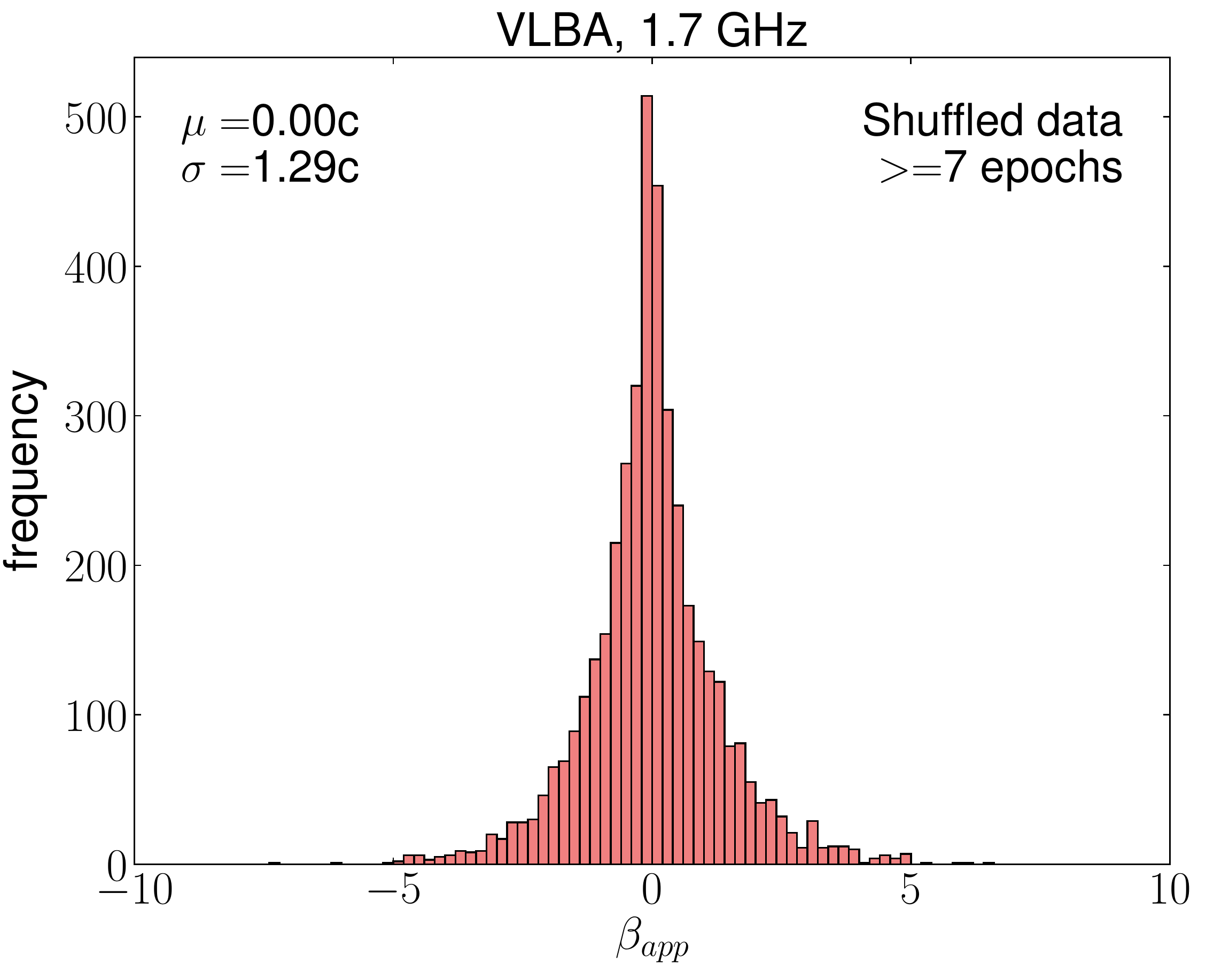}
\caption{Histograms of apparent speeds of vectors identified in more than seven successive epochs by WISE for the original (left) and the hundred shuffled VLBA data (right). The means and standard deviations are noted in the upper left parts.
\label{wiselhisto}}
\end{center}
\end{figure*}

\begin{figure*}[t!]
\begin{center}
\includegraphics[trim=8mm 6mm 21mm 15mm, clip, width = 0.52\textwidth]{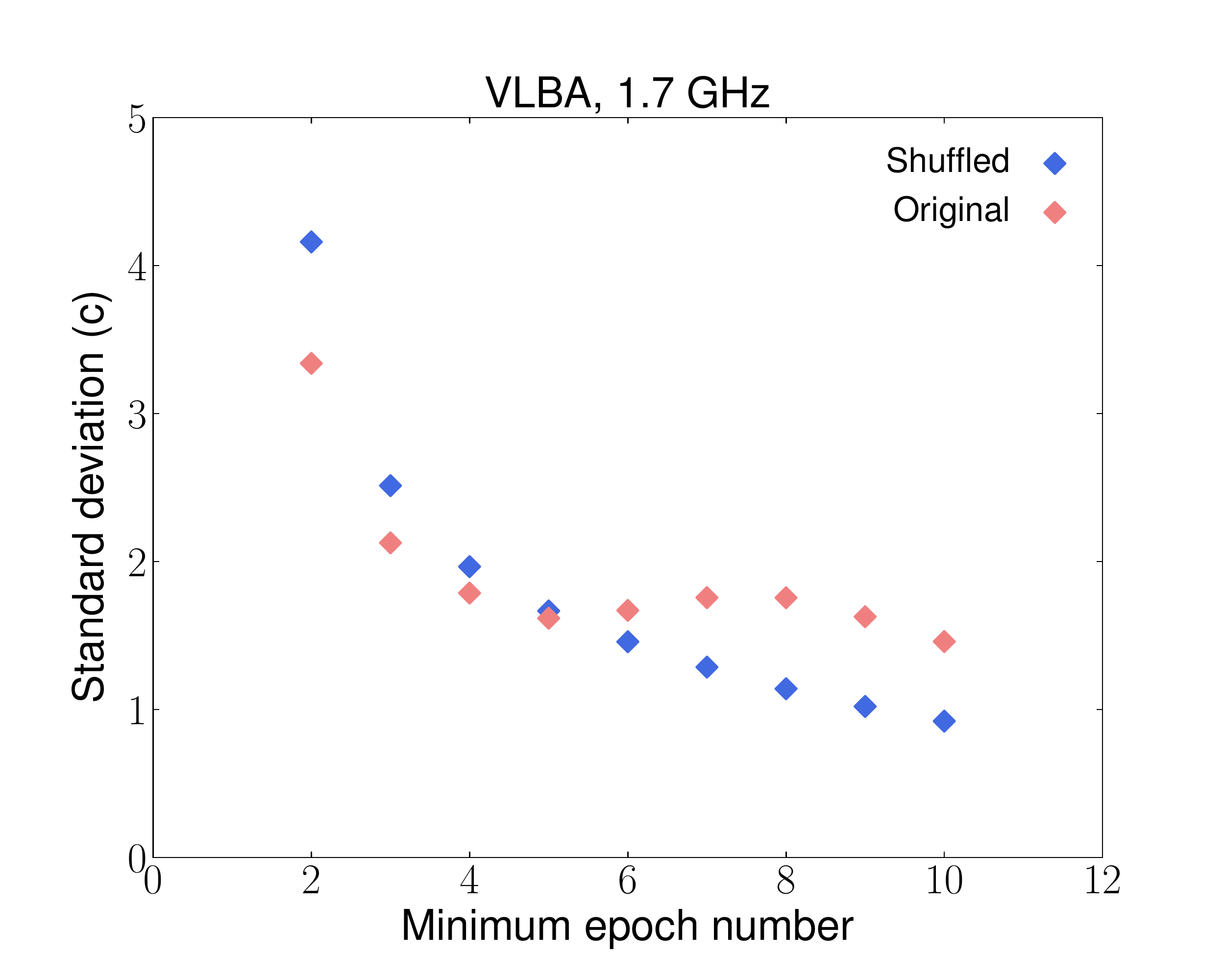}
\caption{Standard deviation of the histograms of apparent speeds detected by WISE for the original (orange filled diamonds) and shuffled VLBA data (blue filled diamonds) as functions of minimum number of successive epochs used for matching SSPs.
\label{wiselepoch}}
\end{center}
\end{figure*}

We perform the WISE analysis for both the original data and one hundred shuffled data again but using a symmetric velocity range constraint of [$-58$, +58] mas/year. Their apparent speed distributions are presented in the right panel of Figure~\ref{wiselsig}. The shuffled data show symmetric distributions of the speeds with respect to zero speeds at all distances, as expected, while the original data preferentially show positive speeds. These trends are also demonstrated by the un-weighted binned data points, showing an acceleration feature for the original data and zero speeds for the shuffled data. The overall trends detected in the original data for symmetric and asymmetric velocity range constraints are in good agreement with each other, although the latter case generally shows higher speeds because any negative velocity vector faster than $-7$ mas/year between adjacent epochs is not captured. In Figure~\ref{wiselhisto}, we present the histograms of the apparent speeds of the original and shuffled data. Their histograms are clearly distinct, suggesting that they follow different probability distribution functions. This indicates that our WISE results capture intrinsic jet motions rather than randomly produced motions due to a chance alignments.

However, it is difficult to quantitatively estimate the probability that at least some of the observed vectors are not intrinsic to the source. One can expect that the probability would depend on the minimum number of successive epochs for matching SSPs. The larger the minimum number is, the less probability for false detection due to a chance alignment. We compare the standard deviations of the histograms of the original and shuffled data for different minimum epoch numbers in Figure~\ref{wiselepoch}. The standard deviation of the shuffled data decreases with increasing epoch number because very fast motions picked up by chance cannot be sustained for a large number of successive epochs. However, the standard deviation of the original data does not decrease substantially for minimum epoch numbers larger than four or five. This test indicates that many of potential random motions picked up by mis-identification in the original data would disappear if we use SSPs matched in more than four or five successive epochs. However, there are still many motions detected in the shuffled data with the standard deviation similar to that observed in the real data for the minimum epoch numbers of four or five. This indicates that one should use a long length-chain such as seven or eight successive epochs, if possible, to obtain robust results. Therefore, we used SSPs matched in more than seven successive epochs for our VLBA data (Section~\ref{sectvlbawise}), for which we have 19 epochs in total. However, we could not use such a long length-chain for our KaVA data observed in only 8 epochs. Thus, we used a length-chain of four successive epochs but used a higher correlation threshold in the MCC analysis to reduce the probability of false detection (Section~\ref{sectwise}).


\begin{figure*}[t!]
\begin{center}
\includegraphics[trim=0mm 0mm 0mm 0mm, clip, width = 0.503\textwidth]{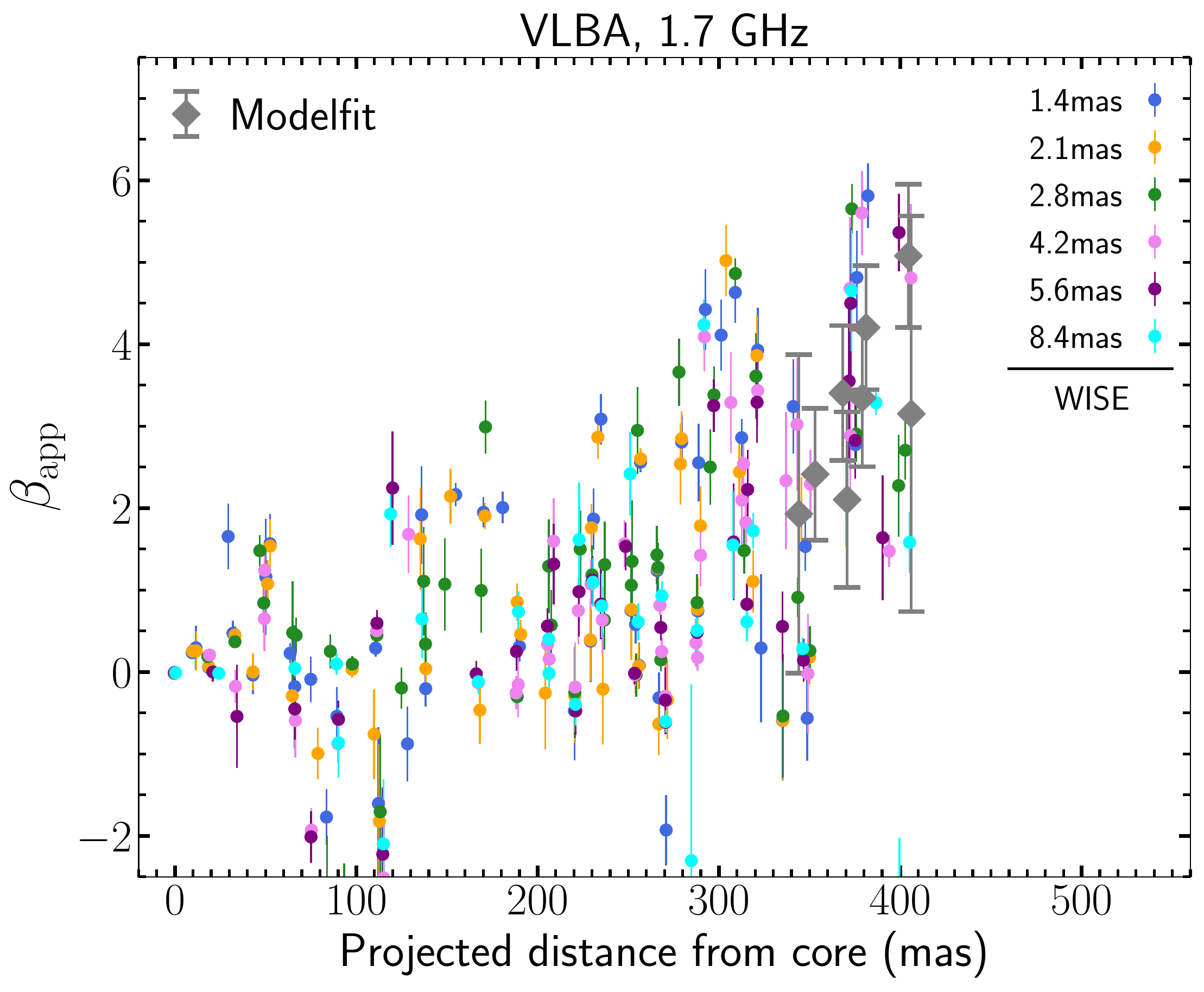}
\includegraphics[trim=0mm 0mm 0mm 0mm, clip, width = 0.488\textwidth]{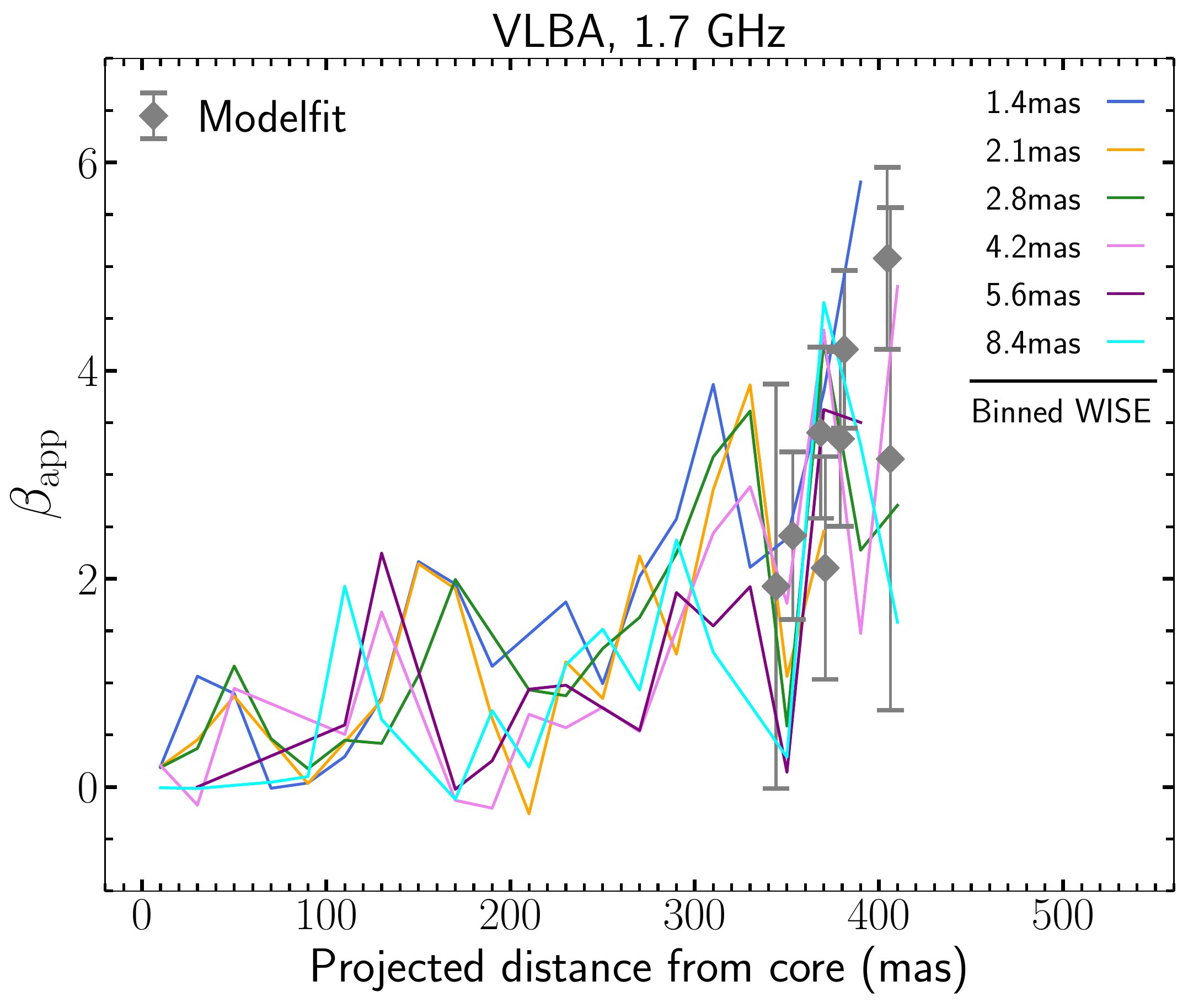}
\caption{Left: apparent speeds in units of the speed of light as functions of projected distance from the core in units of mas detected by the VLBA WISE analysis on six different SWD/IWD scales (shown in different colors), and detected by the {\tt modelfit} analysis (grey filled diamonds). Right: same as the left panel but showing un-weighted binned data points with a bin size of 20 mas.
\label{wiselcomp}}
\end{center}
\end{figure*}

Lastly, we compare the WISE results of the VLBA data on different SWD/IWD scales. The left panel of Figure~\ref{wiselcomp} shows the apparent speeds on different scales as functions of distance. We could observe non-negligible differences in the speeds at several locations but their overall distributions are quite similar to each other. The right panel shows un-weighted binned data points with a bin size of 20 mas, which confirms the similar trends on different scales. The results are also consistent with the {\tt modelfit} results. Thus, we select the results of the finest scale as our representative WISE results for the VLBA data in Section~\ref{sectvlbawise}, which provides the highest effective spatial resolution.

\end{appendix}

\end{document}